\documentclass[%
 reprint,
 amsmath,amssymb,
 aps,nofootinbib,   
]{revtex4-2}
\usepackage{bm}% bold math
\PassOptionsToPackage{linktocpage}{hyperref}
\usepackage[hyperindex,breaklinks]{hyperref}
\usepackage{enumitem}
\usepackage{slashed}

\renewcommand{\theta}{\vartheta}

\renewcommand{\vec}[1]{\ensuremath{\boldsymbol{#1}}}

\newcommand{\N}{\mathbb{N}}	%Symbol for the natural numbers
\newcommand{\bra}[1]{\ensuremath{\left< #1\,\right|}}
\newcommand{\ket}[1]{\ensuremath{\left|\, #1\right>}}

\def\del{\partial}

\usepackage{array}
\usepackage{mathtools}

\usepackage{etoolbox}

\begin{document}

\title{Perturbative Understanding of Non-Perturbative Processes \\
and Quantumization versus Classicalization}

\author{Gia Dvali}
%\email[]{georgi.dvali@physik.uni-muenchen.de}
%\homepage[]{Your web page}
%\thanks{}
%\altaffiliation{}
\author{Lukas Eisemann}
\email[]{lukas.eisemann@physik.lmu.de}

\affiliation{Arnold-Sommerfeld-Center, Ludwig-Maximilians-Universit\"at, \mbox{Theresienstraße 37, 80333 M\"unchen, Germany}}
\affiliation{Max-Planck-Institut f\"ur Physik, F\"ohringer Ring 6, 80805 M\"unchen, Germany}

%\homepage[]{Your web page}
%\thanks{}
%\altaffiliation{}

%\date{\today}

\begin{abstract}

In some instances of study of quantum evolution of 
classical backgrounds it is considered inevitable to resort to non-perturbative methods at the price of treating the system semiclassically.  We show that a fully quantum perturbative 
treatment, in which the background is resolved 
as a multi-particle state, recovers the semiclassical 
non-perturbative results 
and allows going beyond.   We reproduce
particle-creation by a classical field in a theory of two scalars as well as in scalar QED in terms of scattering processes of high multiplicity.
  The multi-particle treatment also
 gives a transparent picture of why 
 a single-process
 transition from a classical to a quantum state, which we call quantumization,  is exponentially suppressed, whereas the opposite process, classicalization, 
 can take place swiftly if the microstate degeneracy of the classical state is high.  An example is provided by the
$N$-graviton portrait of a black hole: a black hole
can form efficiently via a $2\to N$ classicalization process in the collision of 
high-energy particles but its quantumization via a decay $N \to 2$ is exponentially suppressed.

\end{abstract}

%\pacs{(old) 14.60.Pq,13.15.+g,04.60.-m,11.30.Rd}

\maketitle

\tableofcontents

\section{Introduction}\label{sec:intro}

  It is common to 
  classify quantum 
  field theoretic phenomena
into the sub-categories of {\it perturbative}  and 
{\it non-perturbative}  effects. 
A given effect is attributed to the non-perturbative category when
its physical characteristics, such as a cross section or a transition rate, 
are not representable in the form of a perturbative power series in a relevant coupling constant. 

A classic example is provided
 by the Hawking evaporation rate of a black hole.
This rate is proportional to a negative power of the Newtonian gravitational coupling.  It therefore creates the impression that the process is 
non-perturbative in Newton's constant. 
The same reasoning applies to many other examples of particle-creation in a background classical field. 

In reality, in many instances, the underlying physics is fully perturbative
in the coupling.  This perturbativity becomes visible only upon the 
resolution of  the classical background in form of a multi-particle 
state, such as a condensate, which may be realized in form of a number state or a coherent state.  
The corpuscular resolution makes transparent that a seeming non-perturbativity is the result of an interplay between the relevant 
 coupling constant
and the occupation number of quanta. This has been explicitly demonstrated for a number of systems.  

 For example, the theory of  \cite{NPortrait,*BHQPT} offers a resolution of a black hole 
 in form of a condensate (or a coherent state) of gravitons with occupation number $N$. This number is proportional to the inverse of the
gravitational coupling. 
In this picture, Hawking evaporation is described as 
the depletion of the condensate due to re-scattering of constituent gravitons. 
The process is perturbative in Newton's coupling.  
The appearance of
its negative power in the final expression of the evaporation rate is due to the  Bose-enhancement of the process by the combinatorics of the 
occupation number $N$. Since $N$  scales as the inverse of the Newtonian coupling, the rate also comes out to be inversely proportional 
to it.  Hence the apparent non-perturbativity of the process.

 In general, an understanding of non-perturbative 
 mean-field effects as perturbative $N$-particle processes carries obvious
 advantages. In particular, it allows to go beyond the leading order approximation 
 and to capture the $1/N$-corrections coming from quantum effects
 of the individual particles. 
   Such effects are extremely important for properly accounting for the 
   quantum backreaction, which is due to particle-creation on a would-be classical 
   background.
   
 In certain cases this back reaction leads to a complete breakdown of the classical approximation.  
  The concept was originally introduced within the framework of the black hole $N$-portrait  \cite{NPortrait},  with the first explicit simulation 
 of a prototype model  conducted in  \cite{Scrambling},  where the effect was referred to as {\it quantum breaking}.

 Within the black hole $N$-portrait  \cite{NPortrait}, it  has been argued   \cite{BHHair, MacroQ, dS, BHBurden} that 
 quantum breaking takes place at the latest by the time of half-decay.

Beyond black holes, the corpuscular approach to particle-creation 
has been applied to a number of other systems. 
In  \cite{dS, dS1, Dvali:2017eba, dSBurden, dSAnomaly, dSSpecies, dSBRST} the classical de Sitter metric has been resolved as the 
coherent state of gravitons on the Minkowski vacuum.  In particular, such a resolution 
is mandatory within the $S$-matrix formulation of 
gravity, since de Sitter cannot 
serve as a valid $S$-matrix vacuum  \cite{dSAnomaly}.
In this picture, the Gibbons-Hawking particle-creation \cite{Gibbons}
 is described as re-scattering of coherent state gravitons 
into all possible particle species.  
The backreaction due to $1/N$ effects leads to a gradual breakdown of the classical approximation. In particular, this is due to the generation of entanglement  \cite{dS, dS1, Dvali:2017eba, dSAnomaly, dSSpecies}.

 Likewise, an analogous corpuscular study of the decay of the coherently oscillating axion field was given in  \cite{axions}. 
Some further studies on quantum breaking of coherent states 
 using background field methods can be found in \cite{Berezhiani:2020pbv, Berezhiani:2021gph}.
  For other aspects of the corpuscular resolution of de Sitter, 
  cf. \cite{Florian, Florian1, Lasha}. 
 
 In all these examples, the non-perturbative semiclassical effect is recovered as the infinite-$N$ limit of the quantum picture.  
 
 The present paper represents a continuation of the above
 program but 
 with some important novelties that allow to more cleanly 
 extract the specific multi-particle effects in simple examples of basic importance. 
 We study systems that represent the simplest prototype 
 models describing the process of particle-creation by a background field. We do this both in the semiclassical approximation as well as in a fully quantum treatment.
 We show that the perturbative computation
 in the fundamental quantum theory captures 
 the seemingly non-perturbative phenomena obtained in the semiclassical treatment  
 of the same system.
  
  In particular, we make a special focus on the 
 regime when the energies of the produced quanta exceed the oscillation frequency of the background field.  The non-perturbative 
semiclassical analysis gives a very specific suppression of 
particle-creating instabilities.   We show that the fully quantum 
treatment, perturbative in coupling $g^2$, reproduces these
semiclassical results in large-$N$.

  To be more precise, this correspondence becomes exact in the following double-scaling limit, 
\begin{equation}\label{limit}
g^2 \to 0 , \quad \frac{N}{V m^3} \to \infty , \quad g^2 \frac{N}{Vm^3} = \text{fixed} \,,
\end{equation}
where $m$ is the mass of the particles composing 
the would-be classical field and $V$ is the volume.  
However, for finite values of the coupling $g^2$ and the particle number
density $N/V$,
 the fully quantum perturbative analysis 
allows to go beyond the semiclassical approximation and 
capture effects that are higher order in $1/N$.

 We consider three examples.
 In the first example, 
 the quanta of a scalar field $\chi$ are created in the background of an oscillating 
classical scalar field $\phi$.  The coupling between the two fields is $g^2$. 
 At the level of semiclassical analysis,
in which $\phi$ is treated classically, such systems have been widely 
considered and have many applications, e.g., 
for reheating after inflation  \cite{Kofman:1997yn}. 
 In this approximation, the creation of $\chi$-quanta is accounted for by certain instabilities in the background mode equation.  It is usually said that these effects are 
 non-perturbative.  
 This may create the false impression that 
 they cannot be captured by a perturbative analysis.  We show that 
 this is not the case.

 We achieve this by giving a fully quantum resolution of the system.  
 Namely, we represent the classical 
  $\phi$ field as a quantum state of high occupation number $N$. 
  In this language, the particle-creation can be understood as a 
  scattering process in which a number $n$ of $\phi$-quanta 
  is converted into a pair of $\chi$s.  We show that the perturbative treatment 
  fully captures the seemingly non-perturbative effect 
  obtained in the semiclassical theory.

  In two further examples, we generalize the effect to systems with gauge symmetry.  One of these examples considers  production of photons by 
  a time-dependent charged scalar field. Here, too, the semiclassical picture is fully reproduced by perturbative quantum re-scattering of the charged constituent quanta of the condensate. 
  
  The final example is concerned with the inverted situation, i.e., production of a pair of scalar electrons in an oscillating electric field.
In quantum language we describe this process as the creation of a particle-antiparticle pair in the annihilation of many photons. 
These photons represent the quantum constituents of the background electric field.  
  
Our study has implications for fundamental questions regarding 
 classical-to-quantum transitions and vice versa.  
  In particular, it shines light on the question how fast a classical system can transit to a quantum state in which the classical approximation ceases to be  
 valid. In  \cite{Scrambling}, the timescale of such a breakdown 
 was called the {\it quantum break-time}, which we denote by $t_Q$.  The physically relevant timescale to which it must be compared is the characteristic inverse frequency of the constituents of the classical state. In our examples, this is the frequency of the coherent oscillations of the classical field or a condensate.  In the case of the 
$N$-portraits of  black holes  \cite{NPortrait} and of de Sitter  \cite{dS, NPortrait}, 
the basic frequency of the constituent gravitons is given by the inverse of the classical curvature radius.    
 
 The study performed 
 in  \cite{Scrambling}
  shows that in a classically unstable system (i.e., a system with Lyapunov instability), $t_Q$ can scale logarithmically in 
the number $N$ of the system's constituents. Thus, such a system can quantum break swiftly. 
 
 On the other hand, in generic classically-stable or stationary systems, 
 the quantum break-time was argued to be  macroscopic in $N$
  \cite{dS, dS1, Dvali:2017eba}.  
 In such systems the transition to a quantum state is gradual
  and quantum breaking is a cumulative
 effect of a large number of elementary 
processes, each with participation of a small number of quanta.  In particular, this was argued to be the case 
 for black holes  \cite{NPortrait, BHHair, MacroQ, dS, BHBurden},
 as well as for de Sitter  \cite{dS, dS1, Dvali:2017eba, dSBurden, dSAnomaly, dSSpecies}.   
 
  In contrast, the quantum breaking driven by 
   single-process transitions, with participation of order $N$ constituents,  
are expected to be highly suppressed. 
Our analysis contributes to this understanding substantially. Due to the physical importance of this regime, we 
 shall introduce a special term for such a process and
 refer to it as {\it quantumization}.
   
  The flip side of the coin is the process called {\it classicalization}  \cite{Dvali:2010jz}.   
 It represents a process of an inverse transition 
  from a quantum state of few  (say, two) energetic quanta into
  a classical state of soft quanta of high occupation number 
   $N$   \cite{Dvali:2011th, Dvali:2016ovn, Dvali:2014ila, 
   Dvali:2018xoc, Dvali:2020wqi}. 
   Thus, the basis for both
 processes, quantumization as well as classicalization, is the $S$-matrix 
   element between $2$-particle and $N$-particle number eigenstates.  
   
On very general physics grounds 
   one can argue  \cite{Dvali:2020wqi} that the square of 
  such a matrix element is limited from above by an exponentially 
  small factor $e^{-N}$.  This has been confirmed by 
 explicit computations, for example, by the computation 
 of $2\to N$ graviton scattering  \cite{Dvali:2014ila}.
  In fact, for the systems we consider 
  in the present paper, in certain regimes we shall obtain an even  stronger suppression.

Although the basic matrix elements for 
  quantumization and classicalization   processes 
   are the same, their physical 
   manifestations  are very different.   
   The significance of  $2\to N$ versus $N\to 2$
   has already been appreciated within the black hole $N$-portrait
   \cite{NPortrait} and is generic for any classicalizing theory  
 \cite{Dvali:2018xoc, Dvali:2020wqi}.   
     
   The reason is the exponential difference 
   between the degeneracies of $N$-particle and $2$-particle  
   microstates connected by one and the same matrix element. 
    Correspondingly, it makes all the difference whether 
 $N$ appears in the initial or the final state of the process. 
  The classicalization processes, 
 $2 \to N$, can take place with order-one probability provided 
 the $N$-particle classical state has a sufficient 
microstate entropy  \cite{NPortrait, Dvali:2018xoc, Dvali:2020wqi}.  Such states in  \cite{Dvali:2020wqi}
 were called ``saturons", as their entropy is close to saturating the upper bound imposed by unitarity. 
 
 In contrast, the quantumization processes, $N\to 2$, are
 always exponentially suppressed. This is due to the fact that 
 a valid $2$-particle state has insufficient degeneracy for compensating the exponentially suppressed $S$-matrix element.

 A good example of the difference between classicalization 
and quantumization processes is provided by black hole formation 
and decay as described by its $N$-portrait  \cite{NPortrait}.  
Since in this theory a black hole is a state of $N$ soft gravitons,
its formation in a collision of two 
high energy particles represents a  
process of classicalization,  $2\to N$.

The computation of the $2\to N$ graviton process  \cite{Dvali:2014ila, Addazi:2016ksu} shows that the 
 probability of formation for each microstate is 
 $e^{-N}$. However the suppression is compensated 
 by the microstate degeneracy factor $e^{N}$.   
 In other words, the black hole has sufficient degeneracy for compensating  the exponentially suppressed matrix element, 
i.e., it is a saturon  \cite{Dvali:2020wqi}. 
 
 In contrast,  a decay of a black hole into two very energetic quanta,
 represents a quantumization process of the type 
 $N\to 2$.  The microstate degeneracy of the final 
 $2$ particle state is negligible as compared to the exponential suppression.   This explains the well-known property of black holes that, while they can form very fast, they  always decay very slowly. 
 Due to suppressed quantumization
 the decay happens via gradual emission of quanta through Hawking radiation, as opposed to an explosive decay into a few particles.  

Also, one more physical consequence of the suppressed quantumization is an inability of the classical system to
evolve into a highly entangled quantum state via a single-step process. 

The examples studied in the present paper make these generic features of fundamental importance very transparent.

\section{Example 1: $\phi^2\chi^2$}
\label{sec:example1}

For the first example, we consider the model
\begin{equation}\label{L}
L=\frac{1}{2}\partial_\mu\phi \partial^\mu\phi -\frac{1}{2}m^2\phi^2+  \frac{1}{2}\partial_\mu\chi \partial^\mu\chi -\frac{1}{2}m_{\chi}^2\chi^2 - g^2\phi^2\chi^2 \, .
\end{equation}
In a semiclassical treatment, an initial $\phi$-condensate is described by the classical solution
\begin{equation}\label{sol}
\phi_B(t) = \phi_0 \cos \left( m t \right) \,, \quad \chi_B = 0 \, ,
\end{equation}
and the fluctuations around the background \eqref{sol} are quantized:
\begin{equation}\label{fluctuations}
\hat{\phi} = \phi_B + \hat{\delta\phi}   \,, \quad \hat{\chi} = \hat{\delta\chi} \, .
\end{equation}
The initial state we would like to consider is then
\begin{equation}\label{t0Scl}
|t_0\rangle =  |0\rangle_{\delta \phi}  |0\rangle_\chi \, .
\end{equation}
The background alters the propagation of fluctuations to allow for particle-creation out of the vacuum, for example
\begin{equation}\label{sclProcess}
0 \to 2 \chi \, .
\end{equation}
By contrast, in a fully quantum treatment, the condensate is described by an initial state
\begin{equation}\label{initialState}
| t_0 \rangle = | N \rangle_\phi \, | 0 \rangle_\chi \, ,
\end{equation}
where $| N \rangle_\phi$ denotes a state of $\phi$-quanta in the mode of vanishing 3-momentum, $\vec{p}=0$, in a superposition centered around the mean occupation $N$. 
This could, in particular, be a coherent state or simply a number state, and the consideration in the limit \eqref{limit} is independent of the choice. 
For definiteness, we are going to assume a number state 
for now.
The quantum processes giving rise to creation of a pair of $\chi$s are then
\begin{equation}\label{process}
N\phi \to (N-n)\phi + 2\chi \, .
\end{equation}
Here, the final $\phi$s are understood to be also in the $\vec{p}=0$ mode.
Of course there are also processes involving scattered $\phi$s. As compared to \eqref{process}, these are however suppressed by extra powers of the coupling $g^2$ and vanish in the limit \eqref{limit}. Such suppression is not accompanying diagrams with only forward-scattered $\phi$s, but those contribute to the same process \eqref{process} (see also sec.~\ref{sec:parameterRegimes}).
In the absence of an elementary self-interaction of $\phi$, 
and in the regime of negligible Coleman-Weinberg correction to the potential, 
the energetics of the condensate are well approximated by
\begin{equation}\label{NFree}
\frac{N}{V} = \frac{m \phi_0^2}{2 }\, .
\end{equation}
This relates the quantum and classical parameters.
Thus, in semiclassical terms, the limit \eqref{limit} reads 
\begin{equation}\label{limitScl}
g^2 \to 0 \, , \quad \frac{\phi_0^2}{m^2} \to \infty \, , \quad g^2\frac{\phi_0^2}{m^2} =\text{fixed}\,.
\end{equation}
In the following, we are going to compare the quantum-perturbative and semiclassical-non-perturbative prediction for creation of $\chi$-pairs with momenta $k \equiv |\vec{k}|$ corresponding to 
\begin{equation} \label{nRegime}
n \gg \frac{g\phi_0}{m}  \, .
\end{equation}

\subsection{Coherent states and background fields}\label{sec:coherentStates}

Before going to the calculations of the rate, we would like to
recall a general correspondence between calculations in a semiclassical approximation and a fully quantum one that involves coherent states  \cite{Kibble:1965zza}.
In the notation of the example introduced above, consider a coherent state $| c \rangle $ with the property $\langle c | \hat{\phi}(t)|c\rangle = \phi_B(t)$ and $\langle c | \hat{\chi}(t)|c\rangle = \chi_B(t)$, where $\hat{\phi}(t)$ and $\hat{\chi}(t)$ are evolving according to the free Hamiltonian.
For such a state, the S-matrix operator in the presence of a background field agrees with the one of the fully quantum theory between $|c\rangle$,
\begin{equation} \label{correspondence}
\langle c, B | \hat{S}|c, A\rangle_{\phi.\chi} = \langle B |\hat{S}[\phi_B,\chi_B]  | A\rangle_{\delta\phi,\delta\chi} \, ,
\end{equation}  
where the states $|A\rangle$ and $|B\rangle$ denote arbitrary number eigenstates of modes that are not contained in the background.
As we shall see, the convergence of a perturbation series in the field strength depends also on the multiplicity $n$ of the process under consideration.
For the condensate decay \eqref{sclProcess} and \eqref{process} in question, identity \eqref{correspondence} means that in the perturbative regime we have to find agreement of the perturbative fully quantum and non-perturbative semiclassical creation rates with any differences arising from deviations of the initial and final $\phi$-state from $|c\rangle$. Such deviations are model independent and generically vanish in the limit \eqref{limit} as $\sim N^{-1}$.
The following calculations for $\phi^2\chi^2$ as well as those in subsequent sections thus represent explicit examples of the correspondence \eqref{correspondence}.

 \subsection{Quantum calculation}

\begin{figure}
%	\centering 
%	\begin{subfigure}{0.40\textwidth}
\vspace{-0.2\textwidth}
		\includegraphics[width=0.55\textwidth]{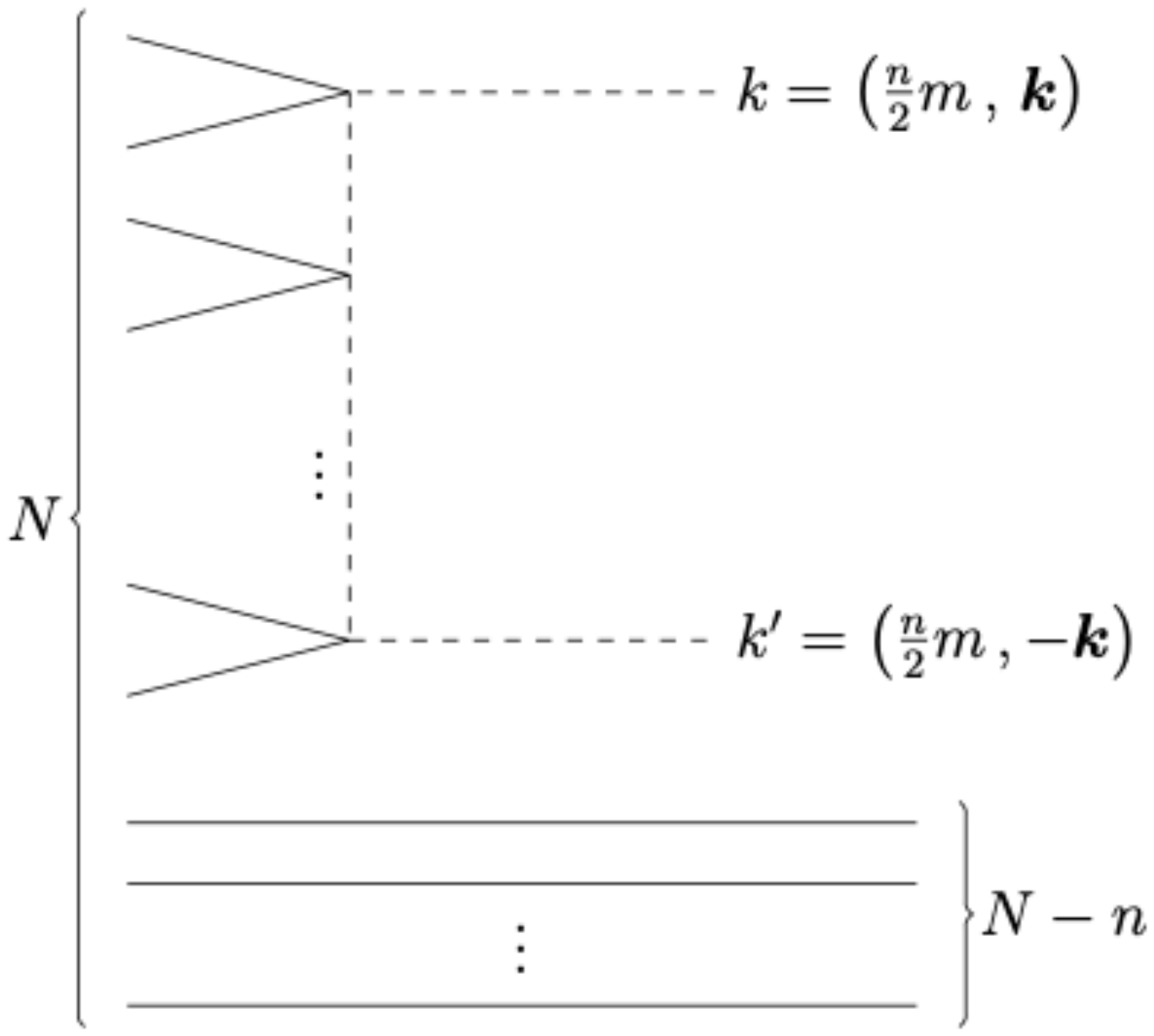}
%	\end{subfigure}
		\vspace{-0.25\textwidth}
	\caption{Diagrammatic representation of the leading order terms in perturbation theory contributing to the amplitude of the process 
	\eqref{process}.
	%$n\phi \to 2\chi$ 
	}
	\label{fig:diagramCrossFour}
\end{figure}

There is only a single diagram (see fig.~\ref{fig:diagramCrossFour}) entering the calculation of the rate for the process \eqref{process} at leading order in $g^2N$.
The number of Wick contractions is canceled by the Taylor-coefficient of the relevant term of the S-matrix operator up to a factor of $2^{n/2}$. The diagram's multiplicity is $n!$ and there is a factor of $1/\sqrt{n!}$ accompanying it because of the initial $n-$fold occupation of the zero mode of $\phi$. Thus the squared amplitude is given by $2^n n! |d|^2$, with $d$ the value of the diagram.
If $q_l$ denotes the the virtual momentum in the propagator following the $l$th insertion of a pair of $\phi$-legs,
then 
$q_l^{\mu} = 2 l m \delta^\mu_0 - k^\mu$ 
and the propagators contribute to $d$ a factor of
\begin{equation}
\prod_{l=1}^{n/2 -1} \left( q^2_l - m_{\chi}^2 \right)^{-1} = 
(-1)^{n/2-1} m^{2-n}\, n^2 \,2^{-n} \, \left( n/2\right)!^{-2} \, .
\end{equation}
Since the initial state does not carry angular momentum, the phase space integration is trivial.
Perturbatively, the kinematics are
\begin{equation}\label{kinematicsQ}
n m =  2 \left( m_\chi^2 + k^2 \right)^{1/2} \,,
\end{equation}
which corresponds to a kinematic threshold for the multiplicity $n$ that is given by 
\begin{equation}\label{thresholdPerturbative}
n> n _0 \equiv \frac{2m_\chi}{m} \,.
\end{equation}
The resulting tree-level rate for $n \phi \to 2 \chi$ is
\begin{equation}\label{Gn}
\Gamma_{n\phi \to 2\chi} = \frac{1}{4\pi} Vm^4 \sqrt{1-\frac{n_0^2}{n^2}}n^4 \left( \frac{g^2}{4 Vm^3} \right)^n   \frac{n!}{(n/2)!^4} \,.
\end{equation}
Perturbation theory can be seen not to break down regardless of the values of $n$.
Including the combinatoric enhancement factor due to the initial $N$ quanta,
\begin{equation}\label{combinatorics}
C_{Nn} = \begin{pmatrix} N \\ n \end{pmatrix} = \frac{N^n}{n!}\left(  1+ \mathcal{O} \left( \frac{n^2}{N}\right) \right)  \,,
\end{equation} 
the rate for the process \eqref{process} to leading order is given by
\begin{equation}\label{rate}
\Gamma \equiv C_{Nn}\Gamma_{n\phi \to 2\chi} \sim \left(  \frac{ e^2}{2 n^2}  \frac{g^2 \phi_0^2}{m^2}\right)^n  \,,
\end{equation}
where we have used relation \eqref{NFree}, the Stirling approximation for the factorial, and omitted factors that scale less strongly with $n$ than exponentially.
The time evolution of $n_k$, the expected occupation number density of $\chi$s per momentum $\vec{k}$, is given for early times by $n_k(t) \sim \Gamma \, t$. 
For later times, however, $\chi$-Bose-enhancement becomes important, leading to an enhanced effective rate
\begin{equation} \label{Bose}
\Gamma_{\text{eff}} \sim \left( 1 + 2 n_k  + \mathcal{O}\left( \frac{n_k^2 n^2}{N} \right)\right)\Gamma \, ,
\end{equation} 
and thus
\begin{equation} \label{realTimeQ}
n_k(t) \sim \exp\left(2 \Gamma  t \right) \,, \quad n_k \gtrsim 1 \, .
\end{equation} 
Therefore, the connection between the quantum rate and the semiclassical prediction is provided by
\begin{equation} \label{exponentQ}
\Gamma \sim \frac{\dot{n}_k}{n_k}  \, .
\end{equation} 
The evolution \eqref{realTimeQ} is of course neglecting modification due to sources of quantum breaking, such as depletion or evolution of entanglement. As long as the number of depleted quanta is much smaller than $N$, it is justified to neglect those effects. 

\subsection{Semiclassical calculation}

In the limit \eqref{limitScl}, the equations of motion simplify. The equation for $\delta \phi$ reads
\begin{align}\label{eomPhi}
\left(\partial^2+m^2\right)\delta\phi &= -2 \left(d_t^2+m^2\right) \phi_B - 2 g^2\chi^2\left(\phi_B + \delta\phi\right) 
\nonumber
\\
&= \mathcal{O}\left( g \right) \,.
\end{align}
The terms involving only the background vanish due to \eqref{sol}.
The equation for $\chi$ reads 
\begin{align}\label{eomChi}
\left(\partial^2+m_\chi^2\right)\chi &= - 2 g^2 \chi \left(\phi_B + \delta\phi\right)^2 
\nonumber
\\
&= -2 g^2 \phi^2_B \chi + \mathcal{O}\left( g \right) \,.
\end{align}
Thus, in the limit \eqref{limitScl}, $\delta \phi$ decouples as a free field and $\chi$ has a time dependent contribution to its mass due to the background.
For the non-perturbative solution, 
we can follow for example \cite{Birrell:1982ix,*Mukhanov:2007zz}.
In the case of a linear equation of the form \eqref{eomChi}, the prediction for $\chi$-creation can be given in real time in terms of the mode function $v_k$, which is defined through the mode expansion of the field operator as
\begin{equation} \label{modeExpansion}
\hat{\chi}(x) = \int d^3k \left( v_{k}(t)\hat{a}_0(\vec{k}) e^{i \vec{k}\cdot \vec{x}} + \textit{h.c.}\right) \,.
\end{equation} 
Here, $\hat{a}_0(\vec{k})$ and its Hermitean conjugate 
are the constant annihilation and creation operators.
The time evolution is contained entirely in the mode function, 
which obeys the equation
\begin{equation} \label{Kmodes}
\left( d_t^2  + \omega^2_k(t)\right) v_k =0 \, ,
\end{equation} 
with
\begin{equation} \label{omegaEff}
\omega^2_k(t) \equiv  m_{\chi}^2 + k^2  + 2  g^2 \phi_0^2 \,  {\rm cos}^2(mt)   \, .
\end{equation}  
The expected particle-creation is then given as
\begin{equation} \label{evol}
n_{k}(t) \equiv \langle 0 | \hat{n}_k (t)| 0\rangle = \frac{1}{2\omega_k} \left(|\dot{v}_k |^2 + \omega^2_k |v_k|^2  \right) -\frac{1}{2} \,,
\end{equation} 
where 
$\hat{n}_k$ is the time evolved operator of the number density of $\chi$-particles per mode $\vec{k}$.
The initial conditions of $v_k$ are constrained by being consistent with the commutation relations of the operators as well as by initially defining the lowest energy state.
Equations \eqref{Kmodes} and \eqref{omegaEff} imply that $v_k$ obeys a Mathieu equation.
In order to find \eqref{evol} in the high multiplicity regime \eqref{nRegime}, %high $n$ (high $k$), 
we can refer to the treatment of the Mathieu equation in  \cite{LL}.
Equations \eqref{Kmodes} and \eqref{omegaEff} can be parametrized as
\begin{equation} \label{Landau}
d_t^2 x + \omega_0^2 \left( 1+ h \cos\left(\gamma t\right)   \right) \, x =0 \, ,
\end{equation} 
where the correspondence of parameters is
\begin{align} \label{par}
\nonumber
\omega_0^2 &\leftrightarrow \overline{\omega_k}^2  \equiv m_{\chi}^2 + k^2 + g^2\phi_0^2 \,,
\\ 
\nonumber
h  &\leftrightarrow \frac{g^2\phi_0^2}{\overline{\omega_k}^2} \,, 
\\
\gamma  &\leftrightarrow 2 m \,.
\end{align} 
Equation \eqref{Landau} can exhibit so called parametric resonance, i.e., has solutions that for certain parameter combinations exhibit exponential growth between the cycles of period $\tau \equiv 2\pi/\gamma$:
\begin{equation} \label{unstableSolution}
x(t+\tau) = e^{s \tau} x(t) \, ,
\end{equation} 
with a parameter of instability $s>0$.
For the particle number one thus has
\begin{equation} \label{solEvol}
\frac{n_{k}(t+\tau)}{n_k(t)} \sim \exp (2 s \tau) \, ,
\end{equation} 
and, coarse graining over several periods, one has
\begin{equation} \label{exponentScl}
\frac{\dot{n}_{k}}{n_k} \sim s \,.
\end{equation} 
In  \cite{LL}, it is shown that for $h \ll 1$, there is parametric resonance in the bands
\begin{equation} \label{kinematicsScl}
\gamma = \frac{2\omega_0}{l} + \epsilon\, , \quad l \in \N \, .
   \end{equation} 
The maximal value of the exponent within these bands scales as
\begin{equation} \label{scalingExponent}
  s \sim h^l  \, ,
\end{equation} 
as does the width $\epsilon$.
Comparing \eqref{kinematicsQ} and \eqref{kinematicsScl}, one has the correspondence
$l \leftrightarrow n/2$.
Therefore, from \eqref{exponentQ} and \eqref{exponentScl}, we see that the quantum rate \eqref{rate} is to be compared with $s$ for $2l$:
\begin{equation} \label{semiclassicalRate}
s \sim h^{2l} \leftrightarrow \left(\frac{g^2\phi_0^2}{m^2} \frac{1}{n^2}\right)^{n} \, ,
\end{equation} 
where we have used \eqref{par} and $h \ll 1$.
The agreement with the parametric scaling of \eqref{rate} can be seen to be complete.

\section{Example 2: Scalar QED}
\label{sec:example2}

A second example is provided by scalar QED with vanishing fundamental self-coupling of the scalar,
\begin{equation}\label{LEx2}
L=D_{\mu}\phi(D^{\mu}\phi)^{\dagger} -m^2\phi^{\dagger}\phi  - \frac{1}{4}F_{\mu \nu}F^{\mu \nu} \, ,
\end{equation}
where $D_\mu \equiv \partial_\mu - i g A_\mu$.
Let us consider the same questions as in the previous example with now $\phi$ and $A$ in the roles of $\phi$ and $\chi$.
That is, semiclassically, we consider out of the vacuum production of a photon pair in the background
\begin{equation}\label{solEx2}
\phi_B = \phi_B^\dagger = \phi_0 \cos (mt) \, , \quad A_B^\mu =0 \, ,
\end{equation}
whereas in the fully quantum picture, we consider the many-particle annihilation processes
\begin{equation}\label{processEx2}
\frac{N}{2} s^- + \frac{N}{2} s^+ \to \frac{N-n}{2} s^- + \frac{N-n}{2} s^+ + 2 \gamma \, .
\end{equation}
quantum and classical parameters are now related by
\begin{equation}\label{NFreeEx2}
\frac{N}{V} = m \phi_0^2\, .
\end{equation}
and the double scaling limit \eqref{limit} in semiclassical terms takes the same form as \eqref{limitScl}.

\subsection{Quantum calculation}

The presence of the 3-point vertex, $i g A^\mu \phi^\dagger \del_\mu \phi + h.c.$, in principle opens up a large variety of Wick contractions contributing to the process \eqref{processEx2} at leading order $\left(N g^2\right)^n$. 
However, all but one of the corresponding diagrams are vanishing due to the combination of gauge redundancy and the special condensate kinematics. This can be seen as follows.
All diagrams involving an incoming scalar pair connected by a single 3-point vertex vanish because the derivative yields a factor of zero in the case of the initial condensate momenta. 
That leaves only diagrams of the type fig.~\ref{fig:diagramCrossFour}, where the dotted lines represent a photon and the 4-point vertices are either elementary or effective ones consisting of two 3-point vertices with one internal $\phi$-line:
\begin{equation}\label{eff4}
-i g^2 \frac{q^\mu_{2l+2}q^\nu_{2l}}{q^2_{2l+1} - m^2} \,.
\end{equation}
Here, $l$ is the number of vertices preceding the vertex. 
In every diagram with one or more vertices like \eqref{eff4}, two momenta $q_l$ are contracted only with photon polarization vectors that are orthogonal:
\begin{equation}
q_l^\mu \, \epsilon_\mu \left( q_{l^\prime}, r \right) = 0 \,, \quad \forall \, \, l,l^\prime \,.
\end{equation}
This relation holds for transverse polarizations $r$, which either belong to an outgoing photon or a neighbouring photon propagator, whose non-transverse part has been projected out by an outgoing photon.
Thus the only non-vanishing diagram is the one constructed solely out of the elementary 4-point vertex, $g^2 A_\mu A^\mu \phi^\dagger \phi$. 
The value of the diagram is different from the 2-scalar case only by a factor
\begin{equation}\label{GnEx2}
\epsilon^*_{\mu}(k,r) \epsilon^{* \mu}(k^\prime,r^\prime) = \delta_{r,r^\prime} \, .
\end{equation}
The resulting tree-level rate for $\frac{n}{2} s^- + \frac{n}{2} s^+ \to 2 \gamma$ for each of the two polarizations is thus identical with \eqref{Gn}.
The combinatoric enhancement factor due to the initial $N$ quanta (using the Stirling approximation) is
\begin{equation}\label{combinatoricsEx2}
C_{Nn} = \begin{pmatrix} N/2 \\ n/2 \end{pmatrix} ^2 \sim 2^{-n} \frac{N^n}{(n/2)!^2}\left(  1+ \mathcal{O} \left( \frac{n^2}{N}\right) \right)  \, .
\end{equation} 
With this and \eqref{NFreeEx2}, the rate for the process \eqref{processEx2} to leading order is given by 
\begin{equation}\label{rateEx2}
\Gamma \equiv C_{Nn}\Gamma_{\frac{n}{2},\frac{n}{2} \to 2 \gamma} \sim \left(  \frac{ e^2}{2 n^2}  \frac{g^2 \phi_0^2}{m^2}\right)^n  \,,
\end{equation}
where again we have used the Stirling approximation and omitted factors that scale less strongly with $n$ than exponentially.

\subsection{Semiclassical calculation}

In the limit \eqref{limitScl}, several terms in the equations of motion are suppressed. 
Furthermore, the background $\phi_B$ obeys a harmonic equation and has vanishing current.
Thus, with the notation
\begin{equation}\label{}
x \equiv \text{Re}\, \delta \phi \, , \quad y \equiv \text{Im}\, \delta \phi  \,,
\end{equation}
the equations for $\delta \phi$ read
\begin{equation}\label{eomX}
\left(\partial^2+m^2\right) x   = \mathcal{O}\left( g \right)
\end{equation}
and
\begin{equation}\label{eomY}
\left(\partial^2+m^2\right) y  = 
g A^\mu \partial_\mu \phi_B + g \partial_\mu \left(A^\mu \phi_B \right) + \mathcal{O}\left( g \right) \,.
\end{equation} 
The equation for $A_\mu$ reads
\begin{equation}\label{eomA}
\partial_\mu F^{\mu\nu} =  
2 g^2 \phi^2_B A^\nu -2 g \left(\phi_B\partial^\nu y - y\partial^\nu \phi_B\right) + \mathcal{O}\left( g \right) \, .
\end{equation} 
One sees that in the limit, $x$ decouples as a free field.
Projecting out the transverse polarizations of $A_\mu$, their equation becomes
\begin{align}\label{eomATransverse}
\left( \partial^2 - 2 g^2 \phi^2_B \right) A^j_T = \mathcal{O}\left( g \right) \, .
\end{align} 
Thus, like in the previous example, the propagating photon degrees of freedom decouple from the fluctuations $\delta \phi$. The creation of $s^+s^-$ is a result of the interaction of $y$ and the Coulomb degree of freedom encoded in $A_0$ and $A^j_L$.
Restricting the attention to processes of photon creation, \eqref{eomATransverse} implies that the mode functions of the two transverse photon polarizations likewise obey a Mathieu equation, 
\begin{equation} \label{KmodesEx2}
\left( d_t^2  + \omega^2_k(t)\right) v_{k,r} =0 \, ,
\end{equation} 
where now
\begin{equation} \label{omegaEffEx2}
\omega^2_k(t) \equiv  k^2  + 2  g^2 \phi_0^2 \,  {\rm cos}^2(mt)   \, .
\end{equation}  
This can again be parametrized as in \eqref{Landau} and the analogous version of \eqref{par}.
Therefore, the semiclassical prediction for the rate is again
\begin{equation} \label{semiclassicalRateEx2}
\frac{\dot{n}_{k,r}}{n_{k,r}} \sim \left(\frac{g^2\phi_0^2}{m^2} \frac{1}{n^2}\right)^{n} \, .
\end{equation} 
Thus, in this example, too, the parametric agreement of \eqref{rateEx2} and \eqref{semiclassicalRateEx2} is complete.

\section{Example 3: Scalar QED with massive photon}
\label{sec:example3}

Let us consider again scalar QED, but now with a non-zero Proca mass $m$:
\begin{equation}\label{LEx3}
L=D_{\mu}\phi(D^{\mu}\phi)^{\dagger} -m_e^2\phi^{\dagger}\phi  - \frac{1}{4}F_{\mu \nu}F^{\mu \nu} +\frac{1}{2}m^2A_\mu A^\mu\, .
\end{equation}
If photon and electron in the preceding example interchange roles, we are dealing with pair-creation in an electric field.
Semiclassically, we are looking at an out of the vacuum creation of $s^+s^-$ in a theory with fields quantized around the background 
\begin{equation}\label{solEx3}
A_B^\mu = \delta^\mu_z \frac{E_0}{m} \cos\left( m t \right) \, , \quad  \phi_B = 0\, .
\end{equation}
The relevant processes in the fully quantum treatment on the other hand are
\begin{equation}\label{processEx3}
N \gamma \to \left(N-n\right)  \gamma + s^+s^- \, .
\end{equation}
quantum and classical parameters are approximately related through
\begin{equation}\label{NFreeEx3}
\frac{N}{V} =  \frac{1}{2m} E_0^2 
\, .
\end{equation}
The double scaling limit \eqref{limit} in semiclassical terms then takes the same form as \eqref{limitScl} with $E_0/m$ in place of $\phi_0$.
Equation \eqref{solEx3} corresponds to a background electromagnetic field of the form
\begin{equation}\label{EMBackground}
F_{0j} = \delta_j^z E_0 \sin \left( \omega t \right)\,, \quad F_{kj} = 0 \,,
\end{equation}
with frequency $\omega = m$.
This field only has an electric component pointing in $z$-direction.
For example, it may serve as an approximate description of the field created in the antinodes of superposing laser light, on length scales short compared to the wavelength, $2\pi/\omega$.
For an optical or X-ray laser, the kinematic threshold is necessarily high, $n_0 \equiv 2m_e/m \gg 1$. This is in contrast to the previous examples with $m_\chi$ arbitrary and $m_\gamma=0$, respectively. 
For such a hierarchy, the dominant process is the one closest to the threshold, 
\begin{equation}\label{thresholdProcess}
n = n_0 + \delta \, ,\quad 0 < \delta \leq 1 \, ,
 \end{equation}
as follows from the scaling of \eqref{rate} and \eqref{rateEx2} and will turn out to be the case here, too.
The regime \eqref{nRegime} for $n \sim n_0$ corresponds to
\begin{equation}\label{nRegimeEx3}
\ \frac{gE_0}{m m_e} \ll 1 \,.
 \end{equation}
 
\subsection{Semiclassical result}
The semiclassical rate of pair-creation in the background field \eqref{EMBackground} averaged over a period of oscillation has been found in 
 \cite{Marinov:1977gq, Brezin:1970xf} 
(see also  \cite{Ringwald:2001ib})
in the regime of $n_0\gg1$ and $E_0 \ll m_e^2/g$.
The full result interpolates between the following two asymptotic expressions.
In the regime of $\frac{gE_0}{m_e\omega} \gg1$, the rate asymptotes to
\begin{align}\label{SchwingerConstantField}
\Gamma \sim &\frac{V m_e^4}{2 \sqrt{2}\pi^4}  \left(\frac{gE_0}{m_e2}\right)^{5/2}\exp\left(-\pi \frac{m_e^2}{gE_0}\right) \,,
\end{align}
which is essentially the suppression obtained by Schwinger 
for the case of a constant electric field \cite{Schwinger:1951nm}.
For the opposite case, $\frac{gE_0}{m_e\omega} \ll1$, the result is asymptotic to 
\begin{align}\label{SchwingerMultiPhoton}
\Gamma \sim \frac{V m_e^4}{(2\pi)^{5/2}}  e^{-2\delta} \left(\frac{\omega}{m_e}\right)^{5/2} 
\left(\frac{e}{4}\frac{g E_0}{m_e \omega}\right)^{2(n_0+\delta)}  \text{Erfi}\left( \sqrt{2\delta} \right) \,.
\end{align}
The latter regime is sometimes referred to as the multi-photon regime and coincides with \eqref{nRegimeEx3}.
In the following perturbative quantum calculation, we indeed fully reproduce \eqref{SchwingerMultiPhoton} in terms of an $n$-photon process \eqref{processEx3}.

\subsection{Quantum calculation}
\begin{figure}
%	\centering 
%	\begin{subfigure}{0.40\textwidth}
		\includegraphics[width=0.4\textwidth]{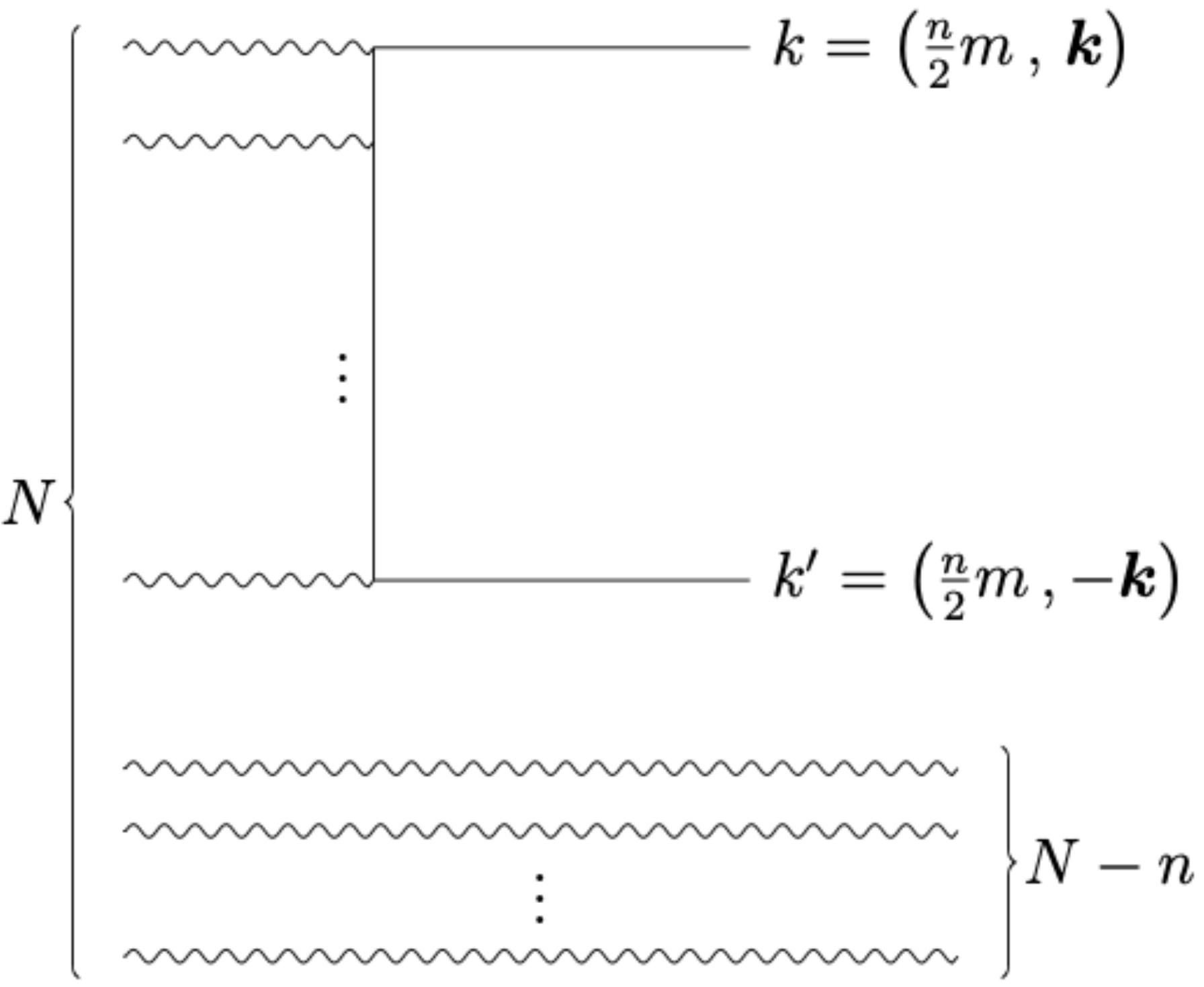}
%	\end{subfigure}
	%	\hspace{0.05\textwidth}
	\caption{One of the terms contributing to the amplitude of the process
	\eqref{processEx3}.
	%$n\phi \to 2\chi$ 
	}
	\label{fig:SchwingerPure3Point}
\end{figure}

In the present case, all possible Wick contractions lead to non-vanishing diagrams. However, near the kinematic threshold, one particular diagram dominates over the others. This is the diagram constructed purely out of the 4-point vertex, as obtained from the digram in fig.~\ref{fig:diagramCrossFour} upon exchanging $\phi$s for $\gamma$s and the $\chi$-pair for $s^+s^-$. 
Let us denote by $\delta \Gamma_4$ the contribution to the rate from the square of only that diagram. Isolating such contributions may be useful despite the contribution of interference terms in the case in which certain diagrams dominate over others.
In terms of the rate \eqref{Gn} in the 2-scalar example, we have
\begin{equation}\label{GnEx3}
\delta\Gamma_4 = 2^{-n-1} \Gamma_{n\phi \to 2\chi} \, .
\end{equation}
Here, the relative factor of $2^{-n}$ comes from the lesser number of Wick contractions since the complex field $\phi$ replaces the real field $\chi$. Likewise, there is another relative factor of $1/2$, since the final particles are not identical. 
All other diagrams
feature photon-insertions via 3-point vertices. The opposite extreme case of the diagram considered above, arising from contracting solely 3-point vertices, is depicted in fig.~\ref{fig:SchwingerPure3Point}.
Let us denote by $\delta\Gamma_3$ the contribution based on its square. It can be obtained in an analogous calculation and the result is
\begin{equation}\label{GnEx3Pure3Point}
\frac{\delta\Gamma_3}{ \delta\Gamma_4} = \frac{e^{2n}}{8n}\left(1-\frac{n_0^2}{n^2}\right)^n \sim \frac{e^{2n}}{8n}\left(\frac{2\delta}{n_0}\right)^{n} \,,
\end{equation}
where the last relation holds for the dominant process \eqref{thresholdProcess}.
This is strongly suppressed for sufficiently small $\delta/n_0$.
The above consideration indicates that a 3-point vertex results in a suppression factor near the threshold and 
that therefore 
for the process \eqref{thresholdProcess} 
the amplitude is well approximated by the diagram in fig.~\ref{fig:diagramCrossFour}.
We thus have
\begin{equation}\label{rateEx3}
\Gamma_{n\gamma \to s^+s^-} = \delta\Gamma_4  \left(1 + \mathcal{O}\left(\delta\right)  \right) \,.
\end{equation}
The combinatoric enhancement is the same as in \eqref{combinatorics}. 
Relating $N$ and $E_0$ via \eqref{NFreeEx3}, we thus have
\begin{equation}\label{GEx3}
\Gamma \sim \frac{1}{8\pi^3} Vm^4 \sqrt{1-\frac{n_0^2}{n^2}}n^2 \left( \frac{e}{4} \frac{g E_0}{m_em}\frac{n_0}{n} \right)^{2n}   \,,
\end{equation}
where we have used \eqref{combinatorics} and the Stirling approximation. 
For the process \eqref{thresholdProcess}, one has $1-(n_0/n)^2 \sim 2\delta/n_0$ as well as $(n_0/n)^{2n}\approx e^{-2\delta}$.
Using these relations, \eqref{GEx3} can be seen to agree completely with \eqref{SchwingerMultiPhoton}, if in the latter one retains only the leading order of
\begin{equation}\label{}
 \text{Erfi}\left( \sqrt{2 \delta} \right)=  2\sqrt{\frac{2}{\pi}}  \sqrt{\delta} \left(1 +  \frac{2}{3}\delta +\mathcal{O}\left(\delta^2\right) \right) \,.
\end{equation}
The relative error for $\delta \ll 1$ is thus $\sim \delta$ and for $\delta = 1$ it is $\approx 0.6$.

\section{Parameter regimes}\label{sec:parameterRegimes}

In the preceding sections, we have considered the rate for a many particle process of the form $n \to 2$ in the example of three different models. 
In a certain regime, we have found a result which scales $\sim \left(g^2N\right)^n$.
In this section, we are going to look in more detail at the limits of validity of this leading order approximation. 
For definiteness, we are going to base the discussion on the $\phi^2 \chi^2$-model. 

\subsection{Semiclassical limit}

In the limit \eqref{limit}, the following two things happen. First, the backreaction from the created 
$\chi$-quanta on the state of $\phi$ vanishes. 
Secondly, loop corrections, that are higher order in $g^2$, vanish as well. 

As a result, any non-perturbative treatment that exploits a semiclassical approximation and neglects radiative corrections becomes exact. 
Of such kind are the semiclassical calculations referenced in the three examples considered.

Their result can be understood as the resummation of the set of diagrams that do not vanish in the limit \eqref{limit}.
Those diagrams are of the same order in $N$ and in $g^2$. At the leading order, 
$n$,
those are the diagrams that we evaluated in the quantum perturbative calculations. At higher orders, $n+k$, those are what may be called forward scattering diagrams, with $k$ $\phi$-quanta all scattered into the initial mode.

As we have seen, in the regime \eqref{nRegime} (regime \eqref{nRegimeEx3}, respectively), which in terms of $N$ reads 
\begin{equation} \label{nRegimeN}
n^2 \gg \frac{g^2 N}{Vm^3} \, ,
\end{equation} 
the diagrams beyond the leading order, $k \geq 1$, are negligible. 
On the other hand, when approaching the opposite regime, the effect of their resummation is both to shift the kinematic threshold through an effective contribution to the particle masses and to alter the scaling of the rate. 
The power-law growth with coupling and field strength gets softened before the rate becomes large.
In the case of the Mathieu analysis (in sec.s \ref{sec:example1} and \ref{sec:example2}), that is the transition between narrow and broad resonance.
For the example of pair-creation in the alternating electric field (sec.~\ref{sec:example3}), the interpolation is shown in \eqref{SchwingerConstantField} and \eqref{SchwingerMultiPhoton}.

\begin{figure}
%	\centering 
%	\begin{subfigure}{0.40\textwidth}
		\includegraphics[width=0.5\textwidth]{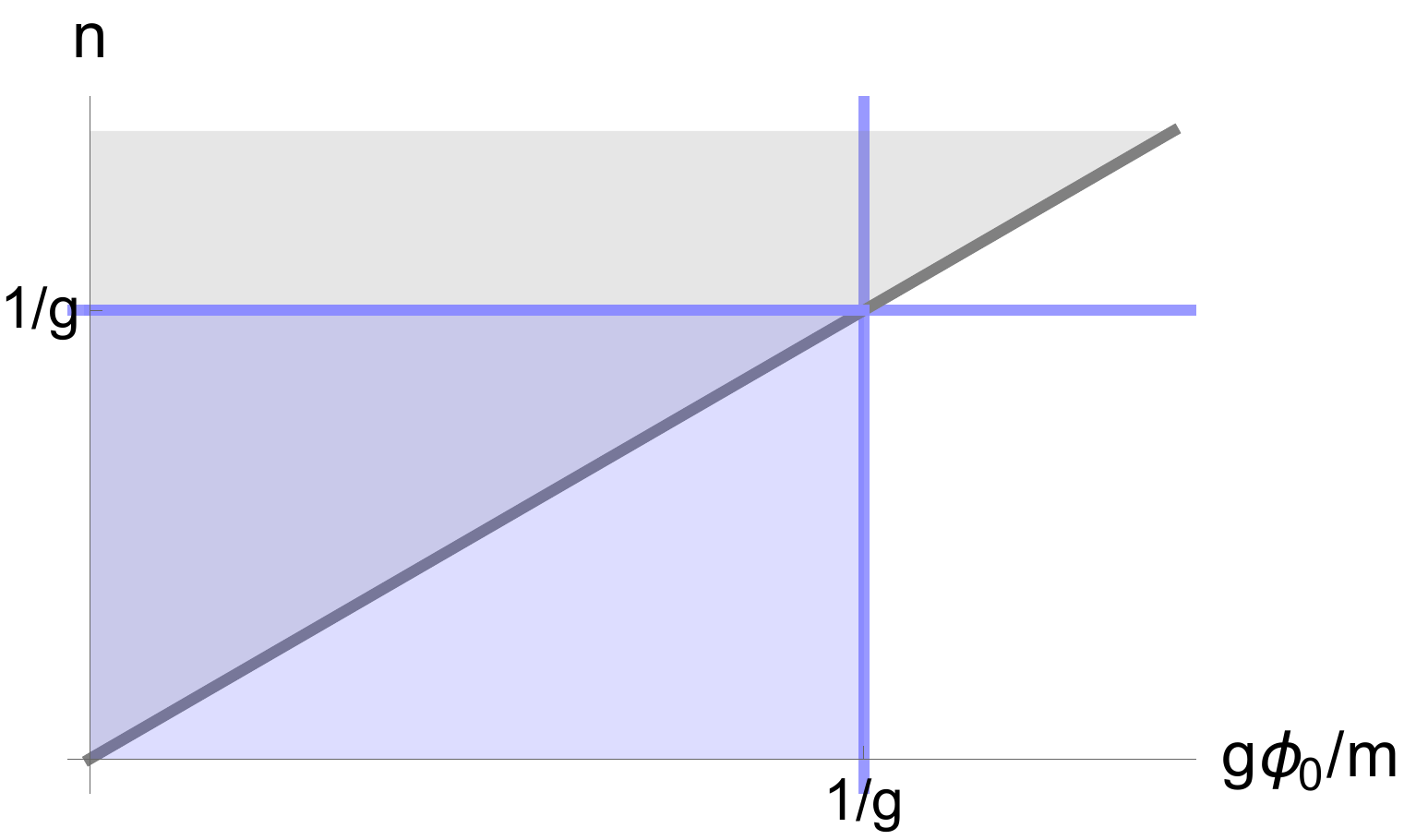}
%	\end{subfigure}
	%	\hspace{0.05\textwidth}
		\caption{Schematic representation of regime boundaries: Above the grey diagonal line, non-perturbative corrections are negligible (see \eqref{nRegimeN}). Additional regimes for finite $g^2$ (and the example of negligible $m_\chi$): To the left of the blue vertical line, loop corrections to the potential are negligible (see \eqref{potentialCrossover}); Below the blue horizontal line, contributions of loop induced diagrams are negligible (see \eqref{nScalingCrossover}). The overlap of coloured areas is the resulting regime in which the calculation is perturbative in both $g^2N$ and $g^2$, i.e., both non-perturbative and quantum corrections to the leading order approximation are negligible.}
	\label{fig:schematicRegimes}
\end{figure}

Fig.~\ref{fig:schematicRegimes} offers a schematic graphical representation of the perturbative and non-perturbative regime in the parameter plane of $n$ and $g \phi_0/m$ as well as further regimes to be discussed below.

\subsection{Finite $g^2$}
\begin{figure}
%	\centering 
%	\begin{subfigure}{0.40\textwidth}
\vspace{-0.15\textwidth}
		\includegraphics[width=0.55\textwidth]{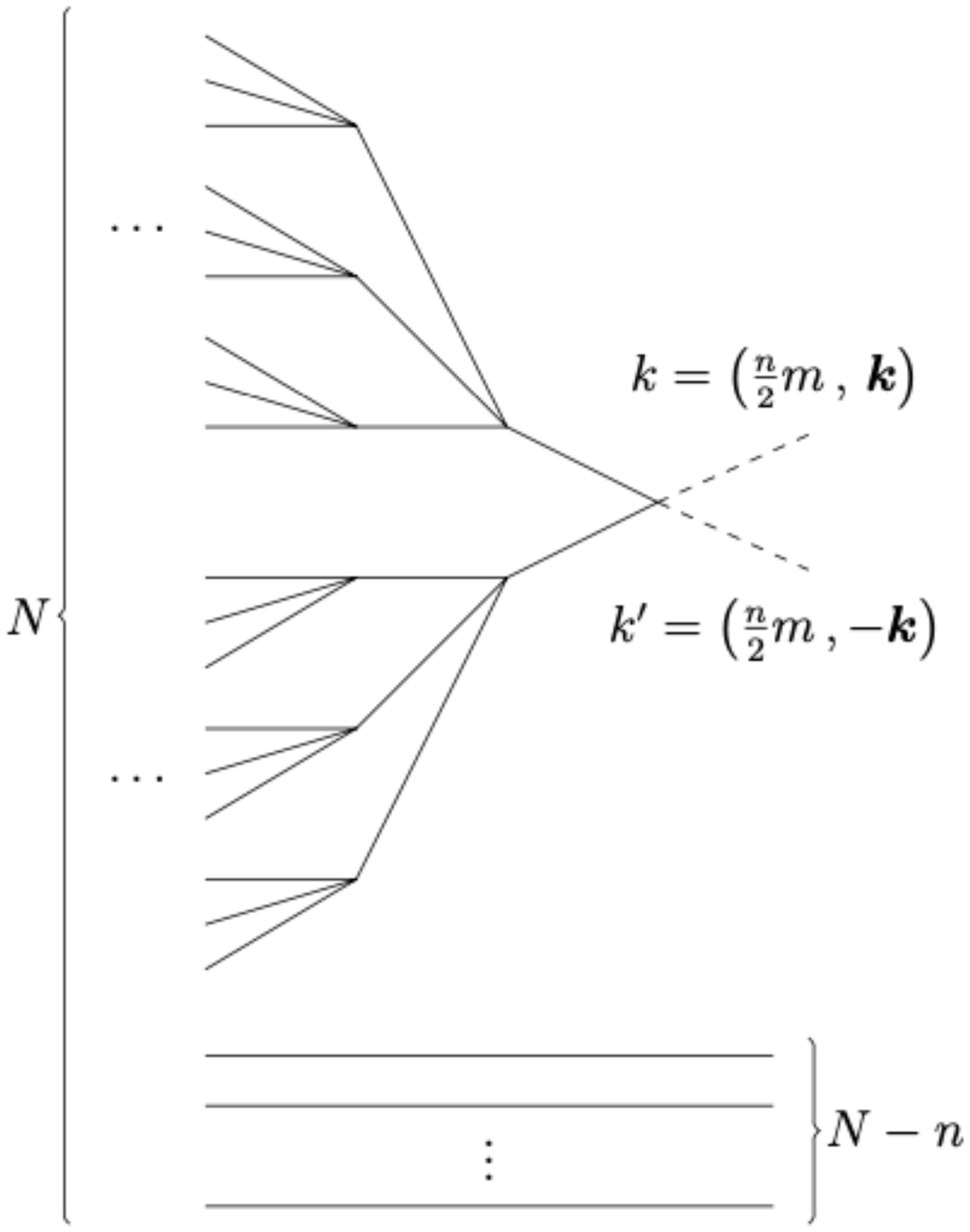}
%	\end{subfigure}
\vspace{-0.17\textwidth}
	%	\hspace{0.05\textwidth}
		\caption{"Symmetrically branching tree" (SBT) diagram: Diagram contributing to the amplitude of the process $n\phi \to 2\chi$ based on only a single cross coupling vertex and otherwise only the quartic self-coupling of $\phi$ (with the most symmetric shape possible).
		 }
	\label{fig:diagramSelf}
\end{figure}
For finite $N$ and $g^2$, there are various quantum corrections arising. 

Corrections of backreaction-type, such as depletion or evolution of entanglement, are $1/N$-suppressed  \cite{dS, dSAnomaly}. Although such corrections build up over time, for the first emissions they are negligible in the case of sufficiently large $N$.

Likewise, corrections due to a single loop or contributions from processes with a single (non-forward) scattered $\phi$ are  
negligible due to the extra power of the coupling $g^2$.

In the following we attempt to identify and estimate other loop effects that may be significant despite $g^2 \ll 1$.
For simplicity, from now on we are going to focus on the case of a negligible mass $m_\chi$.

\subsubsection{Effective potential}
Consider the effective potential at 1-loop \cite{Coleman:1973jx}.
For $\chi = 0$, $\phi$ moves in the potential
\begin{equation}\label{potential}
V_{\textit{1Loop}}(\phi,0)
= \frac{m^2}{2}\phi^2 + \frac{g^4 \phi^4}{16\pi^2}  \left(  \log \left( \frac{2g^2\phi^2}{\mu^2}\right) - \frac{3}{2}\right) \, ,
\end{equation}
where parameters are defined in $\overline{\text{MS}}$.
For $\phi$-values for which the 
loop-correction
is not negligible, the time evolution $\phi(t)$ as in \eqref{sol} 
as well as 
the energetics relating $N$ and $\phi_0$ as in \eqref{NFree} are altered and the considerations of the preceding sections do not go through. 
Equation \eqref{potential} implies that quantum corrections of this kind are only negligible for (with a choice of $\mu \sim m$)
\begin{equation}\label{potentialCrossover}
\frac{\phi_0}{m} \ll \frac{1}{g^2} \, .
\end{equation}
This constrains the initial state in terms of the theory parameter $g^2$ (see fig.~\ref{fig:schematicRegimes}).

\subsubsection{Diagrams with different $n$-scaling}

\begin{figure}
%	\centering 
%	\begin{subfigure}{0.40\textwidth}
\vspace{0.05\textwidth}
		\includegraphics[width=0.25\textwidth]{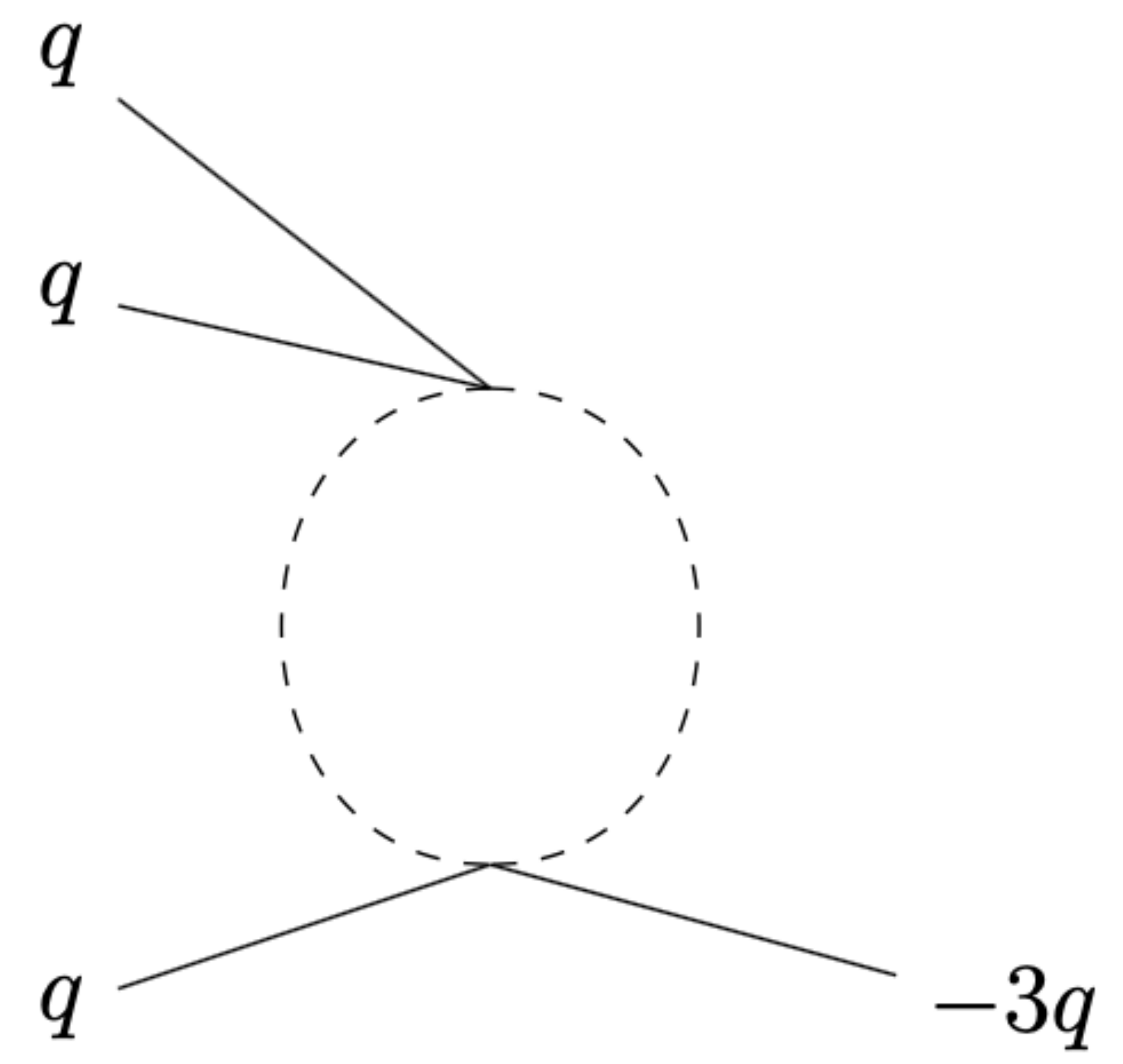}
%	\end{subfigure}
%\vspace{-0.2\textwidth}
	%	\hspace{0.05\textwidth}
		\caption{Momentum dependent $4$-$\phi$-vertex induced by a $\chi$-loop with momenta as occurring in the SBT diagram (fig.~\ref{fig:diagramSelf})
		 }
	\label{fig:chiLoop}
\end{figure}

Another kind of quantum correction may come from loop diagrams that compared to the leading order are suppressed by much higher powers of $g^2$, but on the other hand feature a significantly different scaling with $n$. 
This may happen due to different momentum flow through a diagram permitted by different Wick contractions. In particular, a quartic self-coupling of $\phi$ gives rise to a well-known case of a diagram with strong $n$-scaling  \cite{Goldberg:1990qk, Cornwall:1990hh}. This is the "symmetrically branching tree" (SBT) diagram shown in fig.~\ref{fig:diagramSelf}.

The required self-coupling is induced via the cross-coupling through a $\chi$-loop (see fig.~\ref{fig:chiLoop})
and the corresponding momentum dependent vertex has the value 
\begin{equation}\label{}
v_4 \left(q^2\right) 
= -\frac{i g^4}{16 \pi^2} \left(\log \left( \frac{4q^2}{\mu^2} \right) -2 +i \pi \right) + \mathcal{O}\left( g^6 \right) \, .
\end{equation}
As is evident from fig.s \ref{fig:diagramSelf} and \ref{fig:chiLoop}, we have $m^2 \leq q^2 \leq \frac{1}{9}n^2m^2$, depending on the position of the vertex in the diagram. Therefore, there is no significant $n$-scaling implicit in the momentum dependence of the vertex and 
$4! |v_4| \sim g^4$ 
serves as a good estimate throughout the diagram.

The rate based on only the square of the SBT diagram can be found in a calculation analogous to the ones given in  \cite{Goldberg:1990qk, Cornwall:1990hh}. For the present purposes, the following schematic representation is sufficient:
\begin{equation}\label{Gsbt}
\delta \Gamma_{\textit{SBT}} \sim \left(\frac{c}{Vm^3}\right)^n  n! \,, \quad c \sim g^4 \, .
\end{equation}
Due to the presence of the factorial,
perturbation theory breaks down for
\begin{equation}\label{nMax}
n \gtrsim n_{\textit{max}} \equiv \frac{Vm^3}{c} \, .
\end{equation}
Comparing the contribution \eqref{Gsbt} with the leading order rate \eqref{Gn}, we see that it becomes important for
\begin{equation}\label{sbtCrossover}
n \gtrsim n_{eq} \equiv \sqrt{g^2/c} \, ,
\end{equation}
which takes place well before the breakdown.
There are of course further diagrams as well as interference, but we may take \eqref{sbtCrossover} as an indication for the $n$-regime where contributions due to loop-induced couplings are no longer negligible.
The limit of validity of the leading order approximation \eqref{Gn} is then parametrically given by
\begin{equation}\label{nScalingCrossover}
n \ll \frac{1}{g} \, ,
\end{equation}
which is a bound independent of the initial state parameter $\phi_0$.

As can be seen in fig.~\ref{fig:schematicRegimes}, 
the combination of constraints on $\phi_0$ (on $N$) and $n$ leaves a finite area 
in which corrections due to effects of higher order in $g^2$ and in $g^2 N$ are negligible.

\section{Quantumization and classicalization}%\label{sec:}

 It is an outstanding question under what conditions a classical system  
 can reach a regime in which the classical approximation is 
 no longer applicable and how rapidly this can happen. 
 In this respect, the first important 
   concept, introduced 
  within the
black hole $N$-portrait \cite{NPortrait}, was {\it macro-quantumness} \cite{MacroQ, MacroQ1}. This
term encoded the previous observations that 
$1/N$ quantum effects lead to features that cannot be accounted classically, e.g., emergence of black hole hair
\cite{BHHair}.  
 
The question about the speed of the breakdown of classicality was posed in 
 \cite{Scrambling} (and extended in \cite{dS, Dvali:2017eba}), where  
 this phenomenon was called ``quantum breaking" 
  and the corresponding
 timescale the ``quantum break-time", $t_Q$. 
 
  In  \cite{Scrambling} it was shown that in an $N$-particle system that exhibits a Lyapunov 
 instability, the quantum break-time can be logarithmically 
short, 
\begin{equation} \label{QBT}
t_Q = \lambda^{-1} \ln(N) \,,
\end{equation}
 where $\lambda$ is the Lyapunov exponent.  In particular, this effect was explicitly demonstrated for the
$1+1$-dimensional condensate of bosons on a ring with attractive 
interaction \cite{Scrambling} (for a more recent discussion of this model, see \cite{Kovtun:2020ndc}). 
 
On the other hand, for a system that exhibits no classical instability, 
the situation is different. 
For such a system the following  general bound on the quantum break-time was imposed
 \cite{Dvali:2017eba}:
\begin{equation} \label{tQ}
  t_Q \sim \frac{N}{N^2\Gamma_{2\to2}} \,. 
 \end{equation} 
 Here, $\Gamma_{2\to2}$ is the re-scattering rate of 
 a pair of constituents.  We deliberately kept the $\sim N^2$ factor in the denominator, which comes from the Bose-enhancement of the initial quanta. Basically, it counts the number of pairs in the condensate,  which is $C_{Nn} = N(N-1)/2$. 
 The expression \eqref{tQ} says that in order for the condensate 
 to quantum break, of order $N$ scattering acts must take place.  
 For the example of a self-coupled scalar of \cite{Dvali:2017eba},  
the expression  \eqref{tQ} has been reproduced in  \cite{Berezhiani:2020pbv}
via different methods.

In other words, the main engine for quantum 
breaking is a gradual loss of coherence due to
scattering of small number of constituents into the external quanta,
as opposed to 
non-gradual, i.e., single-process 
transitions with the participation of many constituents. 
For a stable classical system, 
 single-process
 transitions into a quantum state
  are suppressed.

 To this type of processes, in the present paper, we give the special name of {\it quantumization}. Although the term can be defined in
 a broader sense, the processes that we 
 have studied here, specifically amount to transitions from 
an initially classical state into a quantum state with a small number of constituents.  Such processes  
can be viewed as the transitions from macro to micro systems.

  The opposite process, a 
single-process  
  transition from 
  a few-particle quantum state into a classical one, is 
  known as classicalization  \cite{Dvali:2010jz, Dvali:2011th, Dvali:2014ila, Dvali:2016ovn,  Dvali:2018xoc, Dvali:2020wqi}. 
  
  Our analysis contributes into understanding of physics 
  of both processes and in their very different manifestations. 

In particular while unsuppressed classicalization is possible
in certain systems, quantumization 
is universally suppressed.   At first glance, this may sound 
rather surprising, since both processes originate from the 
same basic phenomenon: transition from few to many,  or vice versa. 
 For definiteness, under ``few" let us think of two-particle state 
 $\ket{2}$ and under ``many" the state 
 $\ket{n}$ of  
$n \gg 1$ quanta.  We shall assume that both are approximated as 
valid asymptotic $S$-matrix states.

The basic ingredient controlling both transition 
processes is the square of the $S$-matrix element 
$|\bra{2}\hat{S}\ket{n}|^2$.  This element is always suppressed.
 In fact using very general arguments based on the effective Hamiltonian and locality of the Hilbert space, one can argue 
 that at weak coupling and large $n$ this element is bounded from above (cf. \cite{Dvali:2020wqi}, which refines \cite{Dvali:2018xoc}), 
 \begin{equation} \label{nBound}
 |\bra{2}\hat{S}\ket{n}|^2 \leqslant e^{-n} \,.  
    \end{equation}  
  However, in the estimate of the total probability of the transition, the matrix element 
  is summed over the degeneracy of the final states \cite{NPortrait}, 
  \begin{equation} \label{Total2n}
    \Gamma_{i \to f} \propto \sum_{f} \, |\bra{2}\hat{S}\ket{n}|^2 \,.
  \end{equation}  
 Depending on whether the out state is $\bra{2}$ or $\bra{n}$, 
 the degeneracy factors can be dramatically different. 
 In particular, the degeneracy of two-particle states can never be exponentially large, without compromising the  
 validity of the theory, whereas the degeneracy 
 of $n$-particle states can  \cite{Dvali:2020wqi}. 
  This is the core reason for why quantumization and classicalization 
  are realised in nature very differently. 
  
We shall discuss the two phenomena separately and compare them.
For definiteness, in the following, we are going to refer to the $\phi^2\chi^2$-example. The discussion equally applies to the other examples. 

\subsection{Quantumization} 
 
  The bound \eqref{tQ} provides an explanation of the fact that  
 quantumization phenomena are commonly not observed. 
For example, telecommunication does not suffer from complications due to transitions of electromagnetic waves to mostly quantum states upon encountering a charged particle.
Neither is quantumization expected for objects that are not directly observed. 
To repeat the example given in the introduction, a black hole is not expected to decay into a pair of high energetic photons although no conservation law would bar such a transition.

  The systems studied above allow us to observe the reason for suppression explicitly.  
  In the considered systems, quantumization would correspond 
  to the process $n \rightarrow 2$ with  $n\sim N$.  
 In this case,  an order one fraction of the energy stored in the initial coherent state is transferred to the created particle pair. 

Since the basic $S$-matrix element is suppressed, the 
rate requires enhancement due to combinatorics 
associated with either initial or the final states or both.  

One possibility is the enhancement due to large number 
of choices of $n$ quanta from the total set 
of initial particles. 

However, we must note that the increase of the initial occupation number of quanta up to a level $N \sim g^{-4}$, will invalidate the weak coupling treatment.  Firstly, in such a regime,  the collective interaction from the rest, substantially modifies 
the dispersion relation of each particle.  Secondly, for such 
a high occupation number, the condensate develops Lyapunov instabilities and a 
coherently oscillating field \eqref{sol} no longer 
represents a valid classical background.
This is also clear from the one-loop 
effective potential \eqref{potential}, which corrects the 
free oscillations for amplitudes exceeding $m/g^2$.  
Correspondingly, the validity of our treatment is 
limited 
by \eqref{potentialCrossover}, which in terms of $N$ reads
\begin{equation}\label{boundN}
N \ll \frac{V m^3}{g^4} \, .
\end{equation}

To make the above explicit in the example of a condensate decaying into a particle pair, let us compare the predicted decay time to the timescale of an oscillation, $\sim m^{-1}$, or to  the one required for a complete gradual decay,
\begin{equation}\label{referenceTimeScale}
\tau_{\textit{gradual}} \lesssim N \, \Gamma_{2\to2}^{-1} \sim m^{-1} N \frac{V m^3}{g^4} \, .
\end{equation}
This upper bound is neglecting the Bose-enhancement due to the $N$ quanta in the initial state (which is taken into account in \eqref{tQ}) as well as the accumulating Bose-enhancement due to the created $\chi$-particles (corresponding to the occurrence of parametric resonance in the Mathieu-equation \eqref{Kmodes}). We shall see that nevertheless non-gradual decay is negligible during timescales on the order of \eqref{referenceTimeScale}.

The coherent superposition is sharply peaked around the mean occupation $N$, but it has non-zero support for other occupations, including ones higher than $N$. In order to parametrize the range of $n$ corresponding to $n\sim N$ we may therefore introduce a parameter $r$ and sum the rates for processes with $N/r \leq n \leq r \,N$ (where appropriate rounding is implicit and values $r \gtrsim 10$ are sensible). 
This range of $n$ is in the perturbative regime \eqref{nRegimeN} as long as 
\begin{equation}\label{perturbativeInstantaneous}
r^2 \ll N Vm^3/g^2 \, .
\end{equation}
Depending on the number state in the coherent superposition, for a given $n$, there is a number $n^\prime$ of unscattered initial quanta. 
The sum over final states therefore includes a sum over $n^\prime$. To see what effective combinatoric enhancement that sum amounts to for a given $n$, we note that
\begin{align}\label{}
\nonumber
&\sum_{n^\prime=0}^\infty \left| \langle n^\prime |\hat{a}^n | c \rangle \right|^2 =
\sum_{n^\prime=0}^\infty \begin{pmatrix} n+n^\prime \\ n \end{pmatrix} \left| \langle 0| \hat{a}^n | n \rangle\langle n+n^\prime | c \rangle \right|^2 =
\\
= &\frac{N^n}{n!} \left| \langle 0| \hat{a}^n | n \rangle \right|^2 \, ,
\end{align}
where the occupations of the number states such as $| n \rangle$ and of the coherent state $| c \rangle$ are all referring to the same mode and $\hat{a} |c\rangle = \sqrt{N} |c\rangle$. 
We therefore have for the rate of non-gradual decay
\begin{equation}\label{rateInstantaneous}
\Gamma_{n\sim N} \sim \sum_{n= N/r}^{r\,N} \frac{N^n}{n!} \Gamma_{n \to 2} \, .
\end{equation}
In the regime where for the quantity $\Gamma_{n \to 2}$ the result \eqref{Gn} is valid, we see that the timescale $\Gamma^{-1}_{n\sim N}$ is much larger than \eqref{referenceTimeScale}, as long as \eqref{perturbativeInstantaneous} is fulfilled.
If instead the scaling \eqref{Gsbt} is valid, 
the summation range of $n$ is within the perturbative unitarity bound \eqref{nMax} if $N \leq n_{\textit{max}}/f$, which coincides with \eqref{boundN}. In that case we see that $\Gamma^{-1}_{n\sim N}$ is still much larger than \eqref{referenceTimeScale}.

 Likewise, the rate of quantumization cannot be enhanced significantly through degeneracy of the 
 final $2$-particle states.  The reason is that the exponential 
 increase of $2$-particle degeneracy
  would make the theory strongly coupled, thereby 
 invalidating
  the degrees of freedom. 
  
  A simple way for illustrating this is to notice that the degeneracy 
  of final states can be 
 increased by endowing the $\chi$-particles with an internal 
 ``flavor" quantum number 
 $j=1,2,...,N_f$.   For example,  $\chi_j$ can form an 
 $N_f$-dimensional representation of the $SO(N_f)$ symmetry  group.  This of course will increase the 
rate \eqref{Gn} of the transition $n\phi \to 2\chi$ by a factor $N_f$. 
However, $N_f$ cannot be arbitrarily large. 
 The requirement that  $\phi$ and $\chi$-particles be  
valid (weakly interacting) degrees of freedom, puts the following  
 bound on the collective coupling: 
 \begin{equation} \label{Nfbound}
   N_fg^4 \, \lesssim \, 1\,.
\end{equation}
Violation of this bound changes the regime of the theory. 
In particular this is signalled by the breakdown of the loop expansion. 

 The above bound makes the enhancement
 of the quantumization process negligible as compared 
to the exponential suppression of the matrix element \eqref{nBound}. 

 Notice that the suppression of a fast transition 
 from the classical state of $n$ $\phi$s into a quantum 
 state of two $\chi$s also illustrates the difficulty of 
 rapid generation of quantum entanglement 
 from the classical state. Indeed, the $SO(N_f)$-invariant 
 $2$-particle state, 
  \begin{equation} \label{ENT}
  \ket{2} = \frac{1}{\sqrt{N_f}} \sum_{j=1}^{N_f}
  \ket{\chi_j}\times \ket{\chi_j} \,, 
\end{equation}
is entangled with respect to the flavor quantum number. 
However, its production rate is highly suppressed. 
 Instead in a classically stable system the 
 entanglement can only be generated gradually 
 and requires the time of order \eqref{tQ}  \cite{dS, BHBurden, dSSpecies}.

\subsection{Classicalization}

The story is very different for classicalization 
processes  \cite{Dvali:2010jz}.  
Such a process comes from the inverse transition 
$2 \to n$  \cite{Dvali:2011th, Dvali:2014ila, Dvali:2018xoc, Dvali:2020wqi}. 

Of course, for a fixed pair of initial and final states, the absolute values of 
$S$-matrix elements for transitions 
in both directions, $|S_{2\to n}|$ and $|S_{n\to 2}|$,
are exactly equal.
However, 
the degeneracy of an $n$-particle final state is potentially much 
higher. If the degeneracy is sufficiently high, it can overcome the suppression.  This is 
the essence of classicalization  \cite{Dvali:2018xoc, Dvali:2020wqi}.

 For example, endowing $\phi_j$ with the 
$SO(N_f)$ index, exponentially increases the degeneracy 
of the available states with $n$ $\phi$-bosons
 \cite{Dvali:2020wqi}. 
Of course, by symmetry, the transition elements to each member of the same irreducible representation are equal. The end-result is enhanced 
by the dimensionality of the representation. 
For example, a symmetric state of $n$
quanta  has degeneracy $C_{N_f+n,n} = 
(N_f+n)!/n!N_f!$, which scales as $e^n$ for 
$N_f \sim n$.   Of course,  we do not claim that a simple 
$\phi^2\chi^2$-theory exhibits classicalization
at order-one rate, but the tendency of 
enhancement of classicalizing transition 
$2 \to n$ versus quantumization process $n \to 2$  is clear. 
\footnote{Notice that one could try to enhance the  
 rate of an $n \to 2$ quantumization process by taking
 the initial $n$-particle state as the normalized invariant integral over the entire representation space. This would enhance the amplitude 
 by the dimensionality of the representation. However, such an initial 
 state represents a superposition of all 
 basis state vectors and cannot be regarded as classical. Therefore, 
 such a transition does not fall in our category of 
 quantumization processes. However, such transitions can potentially play an important role in the evolution of a macroscopic system, such as a black hole, after its quantum break-time, since by then 
 the system can evolve into an entangled superposition of many microstates (see the discussion in \ref{sec:outlook}).}   
    
 Correspondingly, the $n$-particle states of high entropy, 
 the so-called saturons,  
 can in principle be produced with unsuppressed rate 
 in a collision of two energetic quanta 
  \cite{Dvali:2020wqi}. 
   Nature provides an example of a saturon in the form of black holes.

It has been believed for a long time  \cite{tHooft, Amati,*AmatiStar, Gross,*GrossStar} 
 that a black hole can form  
in a collision of two particles with unsuppressed probability. 
This view is supported by the semiclassical intuition that 
a black hole must form whenever energy is localized within 
its gravitational radius. 
 In particular, this feature served as the basis for the idea
 of self-completion of gravity \cite{selfC}.
 
However, as usual, without a microscopic theory of a black hole, it is impossible to verify and understand the underlying quantum mechanism for 
its unsuppressed formation.  
 This is provided by the black hole $N$-portrait  \cite{NPortrait}, 
 where a black hole is described  
 as a condensate of $N$ soft gravitons, with 
the  microstate entropy $\sim N$.   

In this theory, the creation of a black hole from colliding quanta represents an instance of classicalization, $2 \to N$, 
in which a highly degenerate $N$-graviton state is formed \cite{NPortrait}. 
 Due to the high multiplicity involved in the process, the probability of transition to each microstate   
is suppressed by $e^{-N}$ \cite{Dvali:2014ila, Addazi:2016ksu}.
Despite this, the total probability of black hole formation 
is of order one.   
The reason for this is found in the non-perturbative enhancement of the transition rate due to 
degeneracy of the $N$-graviton state, 
which accounts for
the black hole entropy  \cite{NPortrait, Dvali:2014ila}.  

  The high multiplicity nature of a black hole 
   likewise serves as an explanation of stability against instantaneous decay.  Indeed, such a decay, $N \to 2$, is controlled by the 
   same exponentially suppressed $S$-matrix element
  as the formation process, $N\to 2$ \cite{NPortrait, Dvali:2014ila}. 
  However, the crucial difference from the formation case is 
  the lack of the entropy-enhancement.  
  This is due to an insufficient degeneracy of 
   the final $2$-particle state. 
   This explains why a classical black hole cannot explode into 
 two (or few) particles.
 
  Of course, the term ``explosion into few particles" in the case of a 
  black hole must be taken with a grain of salt, since the decay products 
  gravitate.  Correspondingly, if the particles materialize  at 
  separations not much larger than the initial gravitational radius, 
  the accompanying classical gravitational field is 
  too strong to be neglected.
  Calculationwise, this merely amounts to 
  a dressing of the outgoing energetic quanta with a classical 
  gravitational field in the form 
  of a coherent state of secondary ``softer" gravitons.

It has been suggested  \cite{Dvali:2021ooc} that another candidate for the  
saturated entropy state is a Color Glass Condensate of gluons in QCD  \cite{CGC}. The formation of this state in collision of protons can then be viewed as a process of classicalization.    
  It has also been proposed  \cite{Dvali:2020wqi} that confinement in QCD with large number of colors can be interpreted as the process 
  of formation of a saturated state of gluons due to excessive 
  color degeneracy. 
  
  The formation
 of saturon bound states 
 in a
 $2 \to N$ process within the Gross-Neveu model \cite{GrossNeveu}
has been studied in \cite{GNsaturons}.
This analysis clearly indicates the entropy enhancement 
of an
otherwise exponentially suppressed transition rate.

\section{Summary and outlook}\label{sec:outlook} 
In the present paper we have  resolved a  class of non-perturbatuve semiclassical phenomena 
in terms of
perturbative quantum processes involving many quanta.  
 This continues the line previously applied 
 to objects such as black holes \cite{NPortrait} and de Sitter
 \cite{NPortrait, dS}  (see section \ref{sec:intro} for a more complete list of references). 
 This resolution makes certain properties very transparent. 
 One very important physical implication is the universal suppression 
 of \textit{quantumization}, defined as a single-process transition from a classical  
 object to a few-particle quantum state. 
 
 Classically stable objects, 
  such as 
 coherently oscillating fields, black holes, or de Sitter space  
 do not ``explode" via creation of few highly energetic quanta. 
 Rather they loose coherence gradually, via quantum processes 
 that involve scattering of a small number of constituents. 
  Likewise, entanglement cannot be generated via a single-process transition but requires a gradual development.

  The fundamental effect unifying all these seemingly   
distinct systems is the suppression of many-particle $S$-matrix elements 
describing $N\to 2$ (or $N \to {\rm few}$) transitions between basic 
number eigenstates.
Previously,  this has been argued to be the case 
in $2\to N$ graviton scattering \cite{NPortrait} and explicitly verified in \cite{Dvali:2014ila}.
The suppression of $2 \to N$ matrix-elements has also been  
argued on very general grounds in \cite{Dvali:2020wqi}.   

 The present study reduces the phenomenon to its bare essentials 
 and explicitly traces its origin  within the most elementary 
 quantum field theories.  Specifically,
 we have explicitly demonstrated the effect within the simplest interacting scalar and gauge theories.

Our analysis is directly extendable to the case of particle-creation in an oscillating background spin-2 field, which, in particular, has been used as a resolution of de Sitter space \cite{dS}.

  We have shown that in the large-$N$ double-scaling 
limit \eqref{limit}, the perturbative quantum results 
 fully match the semiclassical computations, which 
usually are considered to be non-perturbative in nature.  
 
At the same time, our analysis helps to visualize why the phenomenon of classicalization, based on the inverse transition ${\rm few}  \to N$, may be unsuppressed if the $N$-particle state has a sufficient microstate degeneracy \cite{Dvali:2020wqi}.  

In particular, this is the case for black holes \cite{NPortrait}. 
   The microstate degeneracy does not assist the process of 
   quantumization if the initial state of the object
   is described by one particular microstate. 
   Such absence of superposition 
   is a necessary 
   condition of classicality.
   
   It is important that the classicality of a state is 
   distinguished from
   its macroscopic nature. 
  After $t_Q$ the system, e.g., a black hole, continues to be macroscopic but it is no longer classical.  
Due to this, past 
  the half-decay time all bets are off.

   For example, after $t_Q$,  dramatic modifications in black hole 
 evolution can take place. In particular, as shown in  
 \cite{BHBurden}, 
 the backreaction from the stored quantum information (the so-called ``memory burden" effect \cite{Dvali:2018xpy}) 
 becomes overwhelming and the Hawking-approximation breaks down fully. 
 
  Past this point, the following two pathways were outlined 
 in  \cite{BHBurden}. One, as indicated by simulations 
 of prototype systems, is the slowdown of the initially dominant 
 gradual decay process. 
  This, however, does not exclude the development of a new 
classical instability and correspondingly the appearance of a Lyapunov exponent leading to the disintegration of the black hole \cite{BHBurden}.  
Then, in the light of \cite{Scrambling} and of the present discussion, 
this can serve as a trigger of 
unsuppressed quantum processes. 
 That is, at the level of present understanding, 
past its half decay, we cannot exclude a further 
explosive evolution of 
a black hole, neither quantum nor classical.

\acknowledgements
It is a pleasure to thank Max Warkentin for conversations.\\
The figures of diagrams have been created using \textit{TikZ-Feynman} \cite{Ellis:2016jkw}.
\\
This work was supported in part by the Humboldt Foundation under Humboldt Professorship Award, by the
Deutsche Forschungsgemeinschaft (DFG, German Research Foundation) under Germany's Excellence Strategy - EXC-2111 - 390814868, and Germany's Excellence
Strategy under Excellence Cluster Origins.

\bibliography{citations1}

\begin{thebibliography}{53}%
\makeatletter
\providecommand \@ifxundefined [1]{%
 \@ifx{#1\undefined}
}%
\providecommand \@ifnum [1]{%
 \ifnum #1\expandafter \@firstoftwo
 \else \expandafter \@secondoftwo
 \fi
}%
\providecommand \@ifx [1]{%
 \ifx #1\expandafter \@firstoftwo
 \else \expandafter \@secondoftwo
 \fi
}%
\providecommand \natexlab [1]{#1}%
\providecommand \enquote  [1]{``#1''}%
\providecommand \bibnamefont  [1]{#1}%
\providecommand \bibfnamefont [1]{#1}%
\providecommand \citenamefont [1]{#1}%
\providecommand \href@noop [0]{\@secondoftwo}%
\providecommand \href [0]{\begingroup \@sanitize@url \@href}%
\providecommand \@href[1]{\@@startlink{#1}\@@href}%
\providecommand \@@href[1]{\endgroup#1\@@endlink}%
\providecommand \@sanitize@url [0]{\catcode `\\12\catcode `\$12\catcode
  `\&12\catcode `\#12\catcode `\^12\catcode `\_12\catcode `\%12\relax}%
\providecommand \@@startlink[1]{}%
\providecommand \@@endlink[0]{}%
\providecommand \url  [0]{\begingroup\@sanitize@url \@url }%
\providecommand \@url [1]{\endgroup\@href {#1}{\urlprefix }}%
\providecommand \urlprefix  [0]{URL }%
\providecommand \Eprint [0]{\href }%
\providecommand \doibase [0]{https://doi.org/}%
\providecommand \selectlanguage [0]{\@gobble}%
\providecommand \bibinfo  [0]{\@secondoftwo}%
\providecommand \bibfield  [0]{\@secondoftwo}%
\providecommand \translation [1]{[#1]}%
\providecommand \BibitemOpen [0]{}%
\providecommand \bibitemStop [0]{}%
\providecommand \bibitemNoStop [0]{.\EOS\space}%
\providecommand \EOS [0]{\spacefactor3000\relax}%
\providecommand \BibitemShut  [1]{\csname bibitem#1\endcsname}%
\let\auto@bib@innerbib\@empty
%</preamble>
\bibitem [{\citenamefont {Dvali}\ and\ \citenamefont
  {Gomez}(2013{\natexlab{a}})}]{NPortrait}%
  \BibitemOpen
  \bibfield  {author} {\bibinfo {author} {\bibfnamefont {G.}~\bibnamefont
  {Dvali}}\ and\ \bibinfo {author} {\bibfnamefont {C.}~\bibnamefont {Gomez}},\
  }\bibfield  {title} {\bibinfo {title} {{Black Hole's Quantum N-Portrait}},\
  }\href {https://doi.org/10.1002/prop.201300001} {\bibfield  {journal}
  {\bibinfo  {journal} {Fortsch. Phys.}\ }\textbf {\bibinfo {volume} {61}},\
  \bibinfo {pages} {742} (\bibinfo {year} {2013}{\natexlab{a}})},\ \Eprint
  {https://arxiv.org/abs/1112.3359} {arXiv:1112.3359 [hep-th]} \BibitemShut
  {NoStop}%
%%CITATION = ARXIV:1112.3359;%%
\bibitem [{\citenamefont {Dvali}\ and\ \citenamefont
  {Gomez}(2014{\natexlab{a}})}]{BHQPT}%
  \BibitemOpen
  \bibfield  {author} {\bibinfo {author} {\bibfnamefont {G.}~\bibnamefont
  {Dvali}}\ and\ \bibinfo {author} {\bibfnamefont {C.}~\bibnamefont {Gomez}},\
  }\bibfield  {title} {\bibinfo {title} {{Black Holes as Critical Point of
  Quantum Phase Transition}},\ }\href
  {https://doi.org/10.1140/epjc/s10052-014-2752-3} {\bibfield  {journal}
  {\bibinfo  {journal} {Eur. Phys. J.}\ }\textbf {\bibinfo {volume} {C74}},\
  \bibinfo {pages} {2752} (\bibinfo {year} {2014}{\natexlab{a}})},\ \Eprint
  {https://arxiv.org/abs/1207.4059} {arXiv:1207.4059 [hep-th]} \BibitemShut
  {NoStop}%
%%CITATION = ARXIV:1207.4059;%%
\bibitem [{\citenamefont {Dvali}\ \emph {et~al.}(2013)\citenamefont {Dvali},
  \citenamefont {Flassig}, \citenamefont {Gomez}, \citenamefont {Pritzel},\
  and\ \citenamefont {Wintergerst}}]{Scrambling}%
  \BibitemOpen
  \bibfield  {author} {\bibinfo {author} {\bibfnamefont {G.}~\bibnamefont
  {Dvali}}, \bibinfo {author} {\bibfnamefont {D.}~\bibnamefont {Flassig}},
  \bibinfo {author} {\bibfnamefont {C.}~\bibnamefont {Gomez}}, \bibinfo
  {author} {\bibfnamefont {A.}~\bibnamefont {Pritzel}},\ and\ \bibinfo {author}
  {\bibfnamefont {N.}~\bibnamefont {Wintergerst}},\ }\bibfield  {title}
  {\bibinfo {title} {{Scrambling in the Black Hole Portrait}},\ }\href
  {https://doi.org/10.1103/PhysRevD.88.124041} {\bibfield  {journal} {\bibinfo
  {journal} {Phys. Rev.}\ }\textbf {\bibinfo {volume} {D88}},\ \bibinfo {pages}
  {124041} (\bibinfo {year} {2013})},\ \Eprint
  {https://arxiv.org/abs/1307.3458} {arXiv:1307.3458 [hep-th]} \BibitemShut
  {NoStop}%
%%CITATION = ARXIV:1307.3458;%%
\bibitem [{\citenamefont {Dvali}\ and\ \citenamefont
  {Gomez}(2013{\natexlab{b}})}]{BHHair}%
  \BibitemOpen
  \bibfield  {author} {\bibinfo {author} {\bibfnamefont {G.}~\bibnamefont
  {Dvali}}\ and\ \bibinfo {author} {\bibfnamefont {C.}~\bibnamefont {Gomez}},\
  }\bibfield  {title} {\bibinfo {title} {{Black Hole's 1/N Hair}},\ }\href
  {https://doi.org/10.1016/j.physletb.2013.01.020} {\bibfield  {journal}
  {\bibinfo  {journal} {Phys. Lett.}\ }\textbf {\bibinfo {volume} {B719}},\
  \bibinfo {pages} {419} (\bibinfo {year} {2013}{\natexlab{b}})},\ \Eprint
  {https://arxiv.org/abs/1203.6575} {arXiv:1203.6575 [hep-th]} \BibitemShut
  {NoStop}%
%%CITATION = ARXIV:1203.6575;%%
\bibitem [{\citenamefont {Dvali}\ and\ \citenamefont {Gomez}(2012)}]{MacroQ}%
  \BibitemOpen
  \bibfield  {author} {\bibinfo {author} {\bibfnamefont {G.}~\bibnamefont
  {Dvali}}\ and\ \bibinfo {author} {\bibfnamefont {C.}~\bibnamefont {Gomez}},\
  }\bibfield  {title} {\bibinfo {title} {{Black Hole Macro-Quantumness}},\
  }\href@noop {} {\  (\bibinfo {year} {2012})},\ \Eprint
  {https://arxiv.org/abs/1212.0765} {arXiv:1212.0765 [hep-th]} \BibitemShut
  {NoStop}%
%%CITATION = ARXIV:1212.0765;%%
\bibitem [{\citenamefont {Dvali}\ and\ \citenamefont
  {Gomez}(2014{\natexlab{b}})}]{dS}%
  \BibitemOpen
  \bibfield  {author} {\bibinfo {author} {\bibfnamefont {G.}~\bibnamefont
  {Dvali}}\ and\ \bibinfo {author} {\bibfnamefont {C.}~\bibnamefont {Gomez}},\
  }\bibfield  {title} {\bibinfo {title} {{Quantum Compositeness of Gravity:
  Black Holes, AdS and Inflation}},\ }\href
  {https://doi.org/10.1088/1475-7516/2014/01/023} {\bibfield  {journal}
  {\bibinfo  {journal} {JCAP}\ }\textbf {\bibinfo {volume} {1401}},\ \bibinfo
  {pages} {023}},\ \Eprint {https://arxiv.org/abs/1312.4795} {arXiv:1312.4795
  [hep-th]} \BibitemShut {NoStop}%
%%CITATION = ARXIV:1312.4795;%%
\bibitem [{\citenamefont {Dvali}\ \emph {et~al.}(2020)\citenamefont {Dvali},
  \citenamefont {Eisemann}, \citenamefont {Michel},\ and\ \citenamefont
  {Zell}}]{BHBurden}%
  \BibitemOpen
  \bibfield  {author} {\bibinfo {author} {\bibfnamefont {G.}~\bibnamefont
  {Dvali}}, \bibinfo {author} {\bibfnamefont {L.}~\bibnamefont {Eisemann}},
  \bibinfo {author} {\bibfnamefont {M.}~\bibnamefont {Michel}},\ and\ \bibinfo
  {author} {\bibfnamefont {S.}~\bibnamefont {Zell}},\ }\bibfield  {title}
  {\bibinfo {title} {{Black hole metamorphosis and stabilization by memory
  burden}},\ }\href {https://doi.org/10.1103/PhysRevD.102.103523} {\bibfield
  {journal} {\bibinfo  {journal} {Phys. Rev. D}\ }\textbf {\bibinfo {volume}
  {102}},\ \bibinfo {pages} {103523} (\bibinfo {year} {2020})},\ \Eprint
  {https://arxiv.org/abs/2006.00011} {arXiv:2006.00011 [hep-th]} \BibitemShut
  {NoStop}%
\bibitem [{\citenamefont {Dvali}\ and\ \citenamefont {Gomez}(2016)}]{dS1}%
  \BibitemOpen
  \bibfield  {author} {\bibinfo {author} {\bibfnamefont {G.}~\bibnamefont
  {Dvali}}\ and\ \bibinfo {author} {\bibfnamefont {C.}~\bibnamefont {Gomez}},\
  }\bibfield  {title} {\bibinfo {title} {{Quantum Exclusion of Positive
  Cosmological Constant?}},\ }\href {https://doi.org/10.1002/andp.201500216}
  {\bibfield  {journal} {\bibinfo  {journal} {Annalen Phys.}\ }\textbf
  {\bibinfo {volume} {528}},\ \bibinfo {pages} {68} (\bibinfo {year} {2016})},\
  \Eprint {https://arxiv.org/abs/1412.8077} {arXiv:1412.8077 [hep-th]}
  \BibitemShut {NoStop}%
\bibitem [{\citenamefont {Dvali}\ \emph {et~al.}(2017)\citenamefont {Dvali},
  \citenamefont {Gomez},\ and\ \citenamefont {Zell}}]{Dvali:2017eba}%
  \BibitemOpen
  \bibfield  {author} {\bibinfo {author} {\bibfnamefont {G.}~\bibnamefont
  {Dvali}}, \bibinfo {author} {\bibfnamefont {C.}~\bibnamefont {Gomez}},\ and\
  \bibinfo {author} {\bibfnamefont {S.}~\bibnamefont {Zell}},\ }\bibfield
  {title} {\bibinfo {title} {{Quantum Break-Time of de Sitter}},\ }\href
  {https://doi.org/10.1088/1475-7516/2017/06/028} {\bibfield  {journal}
  {\bibinfo  {journal} {JCAP}\ }\textbf {\bibinfo {volume} {1706}},\ \bibinfo
  {pages} {028}},\ \Eprint {https://arxiv.org/abs/1701.08776} {arXiv:1701.08776
  [hep-th]} \BibitemShut {NoStop}%
%%CITATION = ARXIV:1701.08776;%%
\bibitem [{\citenamefont {Dvali}\ \emph {et~al.}(2019)\citenamefont {Dvali},
  \citenamefont {Eisemann}, \citenamefont {Michel},\ and\ \citenamefont
  {Zell}}]{dSBurden}%
  \BibitemOpen
  \bibfield  {author} {\bibinfo {author} {\bibfnamefont {G.}~\bibnamefont
  {Dvali}}, \bibinfo {author} {\bibfnamefont {L.}~\bibnamefont {Eisemann}},
  \bibinfo {author} {\bibfnamefont {M.}~\bibnamefont {Michel}},\ and\ \bibinfo
  {author} {\bibfnamefont {S.}~\bibnamefont {Zell}},\ }\bibfield  {title}
  {\bibinfo {title} {{Universe's Primordial Quantum Memories}},\ }\href
  {https://doi.org/10.1088/1475-7516/2019/03/010} {\bibfield  {journal}
  {\bibinfo  {journal} {JCAP}\ }\textbf {\bibinfo {volume} {1903}},\ \bibinfo
  {pages} {010}},\ \Eprint {https://arxiv.org/abs/1812.08749} {arXiv:1812.08749
  [hep-th]} \BibitemShut {NoStop}%
%%CITATION = ARXIV:1812.08749;%%
\bibitem [{\citenamefont {Dvali}(2020)}]{dSAnomaly}%
  \BibitemOpen
  \bibfield  {author} {\bibinfo {author} {\bibfnamefont {G.}~\bibnamefont
  {Dvali}},\ }\bibfield  {title} {\bibinfo {title} {{$S$-Matrix and Anomaly of
  de Sitter}},\ }\href {https://doi.org/10.3390/sym13010003} {\bibfield
  {journal} {\bibinfo  {journal} {Symmetry}\ }\textbf {\bibinfo {volume}
  {13}},\ \bibinfo {pages} {3} (\bibinfo {year} {2020})},\ \Eprint
  {https://arxiv.org/abs/2012.02133} {arXiv:2012.02133 [hep-th]} \BibitemShut
  {NoStop}%
\bibitem [{\citenamefont {Dvali}(2021{\natexlab{a}})}]{dSSpecies}%
  \BibitemOpen
  \bibfield  {author} {\bibinfo {author} {\bibfnamefont {G.}~\bibnamefont
  {Dvali}},\ }\bibfield  {title} {\bibinfo {title} {Quantum gravity in species
  regime},\ }\href@noop {} {\  (\bibinfo {year} {2021}{\natexlab{a}})},\
  \Eprint {https://arxiv.org/abs/2103.15668} {arXiv:2103.15668 [hep-th]}
  \BibitemShut {NoStop}%
\bibitem [{\citenamefont {Berezhiani}\ \emph
  {et~al.}(2022{\natexlab{a}})\citenamefont {Berezhiani}, \citenamefont
  {Dvali},\ and\ \citenamefont {Sakhelashvili}}]{dSBRST}%
  \BibitemOpen
  \bibfield  {author} {\bibinfo {author} {\bibfnamefont {L.}~\bibnamefont
  {Berezhiani}}, \bibinfo {author} {\bibfnamefont {G.}~\bibnamefont {Dvali}},\
  and\ \bibinfo {author} {\bibfnamefont {O.}~\bibnamefont {Sakhelashvili}},\
  }\bibfield  {title} {\bibinfo {title} {{de Sitter space as a BRST invariant
  coherent state of gravitons}},\ }\href
  {https://doi.org/10.1103/PhysRevD.105.025022} {\bibfield  {journal} {\bibinfo
   {journal} {Phys. Rev. D}\ }\textbf {\bibinfo {volume} {105}},\ \bibinfo
  {pages} {025022} (\bibinfo {year} {2022}{\natexlab{a}})},\ \Eprint
  {https://arxiv.org/abs/2111.12022} {arXiv:2111.12022 [hep-th]} \BibitemShut
  {NoStop}%
\bibitem [{\citenamefont {Gibbons}\ and\ \citenamefont
  {Hawking}(1977)}]{Gibbons}%
  \BibitemOpen
  \bibfield  {author} {\bibinfo {author} {\bibfnamefont {G.~W.}\ \bibnamefont
  {Gibbons}}\ and\ \bibinfo {author} {\bibfnamefont {S.~W.}\ \bibnamefont
  {Hawking}},\ }\bibfield  {title} {\bibinfo {title} {{Cosmological Event
  Horizons, Thermodynamics, and Particle Creation}},\ }\href
  {https://doi.org/10.1103/PhysRevD.15.2738} {\bibfield  {journal} {\bibinfo
  {journal} {Phys. Rev.}\ }\textbf {\bibinfo {volume} {D15}},\ \bibinfo {pages}
  {2738} (\bibinfo {year} {1977})}\BibitemShut {NoStop}%
%%CITATION = PHRVA,D15,2738;%%
\bibitem [{\citenamefont {Dvali}\ and\ \citenamefont {Zell}(2018)}]{axions}%
  \BibitemOpen
  \bibfield  {author} {\bibinfo {author} {\bibfnamefont {G.}~\bibnamefont
  {Dvali}}\ and\ \bibinfo {author} {\bibfnamefont {S.}~\bibnamefont {Zell}},\
  }\bibfield  {title} {\bibinfo {title} {{Classicality and Quantum Break-Time
  for Cosmic Axions}},\ }\href {https://doi.org/10.1088/1475-7516/2018/07/064}
  {\bibfield  {journal} {\bibinfo  {journal} {JCAP}\ }\textbf {\bibinfo
  {volume} {07}},\ \bibinfo {pages} {064}},\ \Eprint
  {https://arxiv.org/abs/1710.00835} {arXiv:1710.00835 [hep-ph]} \BibitemShut
  {NoStop}%
\bibitem [{\citenamefont {Berezhiani}\ and\ \citenamefont
  {Zantedeschi}(2021)}]{Berezhiani:2020pbv}%
  \BibitemOpen
  \bibfield  {author} {\bibinfo {author} {\bibfnamefont {L.}~\bibnamefont
  {Berezhiani}}\ and\ \bibinfo {author} {\bibfnamefont {M.}~\bibnamefont
  {Zantedeschi}},\ }\bibfield  {title} {\bibinfo {title} {{Evolution of
  coherent states as quantum counterpart of classical dynamics}},\ }\href
  {https://doi.org/10.1103/PhysRevD.104.085007} {\bibfield  {journal} {\bibinfo
   {journal} {Phys. Rev. D}\ }\textbf {\bibinfo {volume} {104}},\ \bibinfo
  {pages} {085007} (\bibinfo {year} {2021})},\ \Eprint
  {https://arxiv.org/abs/2011.11229} {arXiv:2011.11229 [hep-th]} \BibitemShut
  {NoStop}%
\bibitem [{\citenamefont {Berezhiani}\ \emph
  {et~al.}(2022{\natexlab{b}})\citenamefont {Berezhiani}, \citenamefont
  {Cintia},\ and\ \citenamefont {Zantedeschi}}]{Berezhiani:2021gph}%
  \BibitemOpen
  \bibfield  {author} {\bibinfo {author} {\bibfnamefont {L.}~\bibnamefont
  {Berezhiani}}, \bibinfo {author} {\bibfnamefont {G.}~\bibnamefont {Cintia}},\
  and\ \bibinfo {author} {\bibfnamefont {M.}~\bibnamefont {Zantedeschi}},\
  }\bibfield  {title} {\bibinfo {title} {{Background-field method and
  initial-time singularity for coherent states}},\ }\href
  {https://doi.org/10.1103/PhysRevD.105.045003} {\bibfield  {journal} {\bibinfo
   {journal} {Phys. Rev. D}\ }\textbf {\bibinfo {volume} {105}},\ \bibinfo
  {pages} {045003} (\bibinfo {year} {2022}{\natexlab{b}})},\ \Eprint
  {https://arxiv.org/abs/2108.13235} {arXiv:2108.13235 [hep-th]} \BibitemShut
  {NoStop}%
\bibitem [{\citenamefont {K\"uhnel}(2015)}]{Florian}%
  \BibitemOpen
  \bibfield  {author} {\bibinfo {author} {\bibfnamefont {F.}~\bibnamefont
  {K\"uhnel}},\ }\bibfield  {title} {\bibinfo {title} {{Thoughts on the Vacuum
  Energy in the Quantum N-Portrait}},\ }\href
  {https://doi.org/10.1142/S0217732315501977} {\bibfield  {journal} {\bibinfo
  {journal} {Mod. Phys. Lett. A}\ }\textbf {\bibinfo {volume} {30}},\ \bibinfo
  {pages} {1550197} (\bibinfo {year} {2015})},\ \Eprint
  {https://arxiv.org/abs/1408.5897} {arXiv:1408.5897 [gr-qc]} \BibitemShut
  {NoStop}%
\bibitem [{\citenamefont {Kuhnel}\ and\ \citenamefont
  {Sandstad}(2015)}]{Florian1}%
  \BibitemOpen
  \bibfield  {author} {\bibinfo {author} {\bibfnamefont {F.}~\bibnamefont
  {Kuhnel}}\ and\ \bibinfo {author} {\bibfnamefont {M.}~\bibnamefont
  {Sandstad}},\ }\bibfield  {title} {\bibinfo {title} {{Corpuscular
  Consideration of Eternal Inflation}},\ }\href
  {https://doi.org/10.1140/epjc/s10052-015-3736-7} {\bibfield  {journal}
  {\bibinfo  {journal} {Eur. Phys. J. C}\ }\textbf {\bibinfo {volume} {75}},\
  \bibinfo {pages} {505} (\bibinfo {year} {2015})},\ \Eprint
  {https://arxiv.org/abs/1504.02377} {arXiv:1504.02377 [gr-qc]} \BibitemShut
  {NoStop}%
\bibitem [{\citenamefont {Berezhiani}(2017)}]{Lasha}%
  \BibitemOpen
  \bibfield  {author} {\bibinfo {author} {\bibfnamefont {L.}~\bibnamefont
  {Berezhiani}},\ }\bibfield  {title} {\bibinfo {title} {{On Corpuscular Theory
  of Inflation}},\ }\href {https://doi.org/10.1140/epjc/s10052-017-4672-5}
  {\bibfield  {journal} {\bibinfo  {journal} {Eur. Phys. J. C}\ }\textbf
  {\bibinfo {volume} {77}},\ \bibinfo {pages} {106} (\bibinfo {year} {2017})},\
  \Eprint {https://arxiv.org/abs/1610.08433} {arXiv:1610.08433 [hep-th]}
  \BibitemShut {NoStop}%
\bibitem [{\citenamefont {Kofman}\ \emph {et~al.}(1997)\citenamefont {Kofman},
  \citenamefont {Linde},\ and\ \citenamefont {Starobinsky}}]{Kofman:1997yn}%
  \BibitemOpen
  \bibfield  {author} {\bibinfo {author} {\bibfnamefont {L.}~\bibnamefont
  {Kofman}}, \bibinfo {author} {\bibfnamefont {A.~D.}\ \bibnamefont {Linde}},\
  and\ \bibinfo {author} {\bibfnamefont {A.~A.}\ \bibnamefont {Starobinsky}},\
  }\bibfield  {title} {\bibinfo {title} {{Towards the theory of reheating after
  inflation}},\ }\href {https://doi.org/10.1103/PhysRevD.56.3258} {\bibfield
  {journal} {\bibinfo  {journal} {Phys. Rev. D}\ }\textbf {\bibinfo {volume}
  {56}},\ \bibinfo {pages} {3258} (\bibinfo {year} {1997})},\ \Eprint
  {https://arxiv.org/abs/hep-ph/9704452} {arXiv:hep-ph/9704452} \BibitemShut
  {NoStop}%
\bibitem [{\citenamefont {Dvali}\ \emph
  {et~al.}(2011{\natexlab{a}})\citenamefont {Dvali}, \citenamefont {Giudice},
  \citenamefont {Gomez},\ and\ \citenamefont {Kehagias}}]{Dvali:2010jz}%
  \BibitemOpen
  \bibfield  {author} {\bibinfo {author} {\bibfnamefont {G.}~\bibnamefont
  {Dvali}}, \bibinfo {author} {\bibfnamefont {G.~F.}\ \bibnamefont {Giudice}},
  \bibinfo {author} {\bibfnamefont {C.}~\bibnamefont {Gomez}},\ and\ \bibinfo
  {author} {\bibfnamefont {A.}~\bibnamefont {Kehagias}},\ }\bibfield  {title}
  {\bibinfo {title} {{UV-Completion by Classicalization}},\ }\href
  {https://doi.org/10.1007/JHEP08(2011)108} {\bibfield  {journal} {\bibinfo
  {journal} {JHEP}\ }\textbf {\bibinfo {volume} {08}},\ \bibinfo {pages}
  {108}},\ \Eprint {https://arxiv.org/abs/1010.1415} {arXiv:1010.1415 [hep-ph]}
  \BibitemShut {NoStop}%
\bibitem [{\citenamefont {Dvali}\ \emph
  {et~al.}(2011{\natexlab{b}})\citenamefont {Dvali}, \citenamefont {Gomez},\
  and\ \citenamefont {Kehagias}}]{Dvali:2011th}%
  \BibitemOpen
  \bibfield  {author} {\bibinfo {author} {\bibfnamefont {G.}~\bibnamefont
  {Dvali}}, \bibinfo {author} {\bibfnamefont {C.}~\bibnamefont {Gomez}},\ and\
  \bibinfo {author} {\bibfnamefont {A.}~\bibnamefont {Kehagias}},\ }\bibfield
  {title} {\bibinfo {title} {{Classicalization of Gravitons and Goldstones}},\
  }\href {https://doi.org/10.1007/JHEP11(2011)070} {\bibfield  {journal}
  {\bibinfo  {journal} {JHEP}\ }\textbf {\bibinfo {volume} {11}},\ \bibinfo
  {pages} {070}},\ \Eprint {https://arxiv.org/abs/1103.5963} {arXiv:1103.5963
  [hep-th]} \BibitemShut {NoStop}%
\bibitem [{\citenamefont {Dvali}(2017)}]{Dvali:2016ovn}%
  \BibitemOpen
  \bibfield  {author} {\bibinfo {author} {\bibfnamefont {G.}~\bibnamefont
  {Dvali}},\ }\bibfield  {title} {\bibinfo {title} {{Strong Coupling and
  Classicalization}},\ }\href {https://doi.org/10.1142/9789813208292_0005}
  {\bibfield  {journal} {\bibinfo  {journal} {Subnucl. Ser.}\ }\textbf
  {\bibinfo {volume} {53}},\ \bibinfo {pages} {189} (\bibinfo {year} {2017})},\
  \Eprint {https://arxiv.org/abs/1607.07422} {arXiv:1607.07422 [hep-th]}
  \BibitemShut {NoStop}%
\bibitem [{\citenamefont {Dvali}\ \emph {et~al.}(2015)\citenamefont {Dvali},
  \citenamefont {Gomez}, \citenamefont {Isermann}, \citenamefont {L\"ust},\
  and\ \citenamefont {Stieberger}}]{Dvali:2014ila}%
  \BibitemOpen
  \bibfield  {author} {\bibinfo {author} {\bibfnamefont {G.}~\bibnamefont
  {Dvali}}, \bibinfo {author} {\bibfnamefont {C.}~\bibnamefont {Gomez}},
  \bibinfo {author} {\bibfnamefont {R.~S.}\ \bibnamefont {Isermann}}, \bibinfo
  {author} {\bibfnamefont {D.}~\bibnamefont {L\"ust}},\ and\ \bibinfo {author}
  {\bibfnamefont {S.}~\bibnamefont {Stieberger}},\ }\bibfield  {title}
  {\bibinfo {title} {{Black hole formation and classicalization in
  ultra-Planckian 2\textrightarrow{}N scattering}},\ }\href
  {https://doi.org/10.1016/j.nuclphysb.2015.02.004} {\bibfield  {journal}
  {\bibinfo  {journal} {Nucl. Phys. B}\ }\textbf {\bibinfo {volume} {893}},\
  \bibinfo {pages} {187} (\bibinfo {year} {2015})},\ \Eprint
  {https://arxiv.org/abs/1409.7405} {arXiv:1409.7405 [hep-th]} \BibitemShut
  {NoStop}%
\bibitem [{\citenamefont {Dvali}(2018{\natexlab{a}})}]{Dvali:2018xoc}%
  \BibitemOpen
  \bibfield  {author} {\bibinfo {author} {\bibfnamefont {G.}~\bibnamefont
  {Dvali}},\ }\bibfield  {title} {\bibinfo {title} {{Classicalization Clearly:
  Quantum Transition into States of Maximal Memory Storage Capacity}},\
  }\href@noop {} {\  (\bibinfo {year} {2018}{\natexlab{a}})},\ \Eprint
  {https://arxiv.org/abs/1804.06154} {arXiv:1804.06154 [hep-th]} \BibitemShut
  {NoStop}%
%%CITATION = ARXIV:1804.06154;%%
\bibitem [{\citenamefont {Dvali}(2021{\natexlab{b}})}]{Dvali:2020wqi}%
  \BibitemOpen
  \bibfield  {author} {\bibinfo {author} {\bibfnamefont {G.}~\bibnamefont
  {Dvali}},\ }\bibfield  {title} {\bibinfo {title} {{Entropy Bound and
  Unitarity of Scattering Amplitudes}},\ }\href
  {https://doi.org/10.1007/JHEP03(2021)126} {\bibfield  {journal} {\bibinfo
  {journal} {JHEP}\ }\textbf {\bibinfo {volume} {03}},\ \bibinfo {pages}
  {126}},\ \Eprint {https://arxiv.org/abs/2003.05546} {arXiv:2003.05546
  [hep-th]} \BibitemShut {NoStop}%
\bibitem [{\citenamefont {Addazi}\ \emph {et~al.}(2017)\citenamefont {Addazi},
  \citenamefont {Bianchi},\ and\ \citenamefont {Veneziano}}]{Addazi:2016ksu}%
  \BibitemOpen
  \bibfield  {author} {\bibinfo {author} {\bibfnamefont {A.}~\bibnamefont
  {Addazi}}, \bibinfo {author} {\bibfnamefont {M.}~\bibnamefont {Bianchi}},\
  and\ \bibinfo {author} {\bibfnamefont {G.}~\bibnamefont {Veneziano}},\
  }\bibfield  {title} {\bibinfo {title} {{Glimpses of black hole
  formation/evaporation in highly inelastic, ultra-planckian string
  collisions}},\ }\href {https://doi.org/10.1007/JHEP02(2017)111} {\bibfield
  {journal} {\bibinfo  {journal} {JHEP}\ }\textbf {\bibinfo {volume} {02}},\
  \bibinfo {pages} {111}},\ \Eprint {https://arxiv.org/abs/1611.03643}
  {arXiv:1611.03643 [hep-th]} \BibitemShut {NoStop}%
\bibitem [{\citenamefont {Kibble}(1965)}]{Kibble:1965zza}%
  \BibitemOpen
  \bibfield  {author} {\bibinfo {author} {\bibfnamefont {T.~W.~B.}\
  \bibnamefont {Kibble}},\ }\bibfield  {title} {\bibinfo {title} {{Frequency
  Shift in High-Intensity Compton Scattering}},\ }\href
  {https://doi.org/10.1103/PhysRev.138.B740} {\bibfield  {journal} {\bibinfo
  {journal} {Phys. Rev.}\ }\textbf {\bibinfo {volume} {138}},\ \bibinfo {pages}
  {B740} (\bibinfo {year} {1965})}\BibitemShut {NoStop}%
\bibitem [{\citenamefont {Birrell}\ and\ \citenamefont
  {Davies}(1984)}]{Birrell:1982ix}%
  \BibitemOpen
  \bibfield  {author} {\bibinfo {author} {\bibfnamefont {N.~D.}\ \bibnamefont
  {Birrell}}\ and\ \bibinfo {author} {\bibfnamefont {P.~C.~W.}\ \bibnamefont
  {Davies}},\ }\href {https://doi.org/10.1017/CBO9780511622632} {\emph
  {\bibinfo {title} {{Quantum Fields in Curved Space}}}},\ Cambridge Monographs
  on Mathematical Physics\ (\bibinfo  {publisher} {Cambridge Univ. Press},\
  \bibinfo {address} {Cambridge, UK},\ \bibinfo {year} {1984})\BibitemShut
  {NoStop}%
\bibitem [{\citenamefont {Mukhanov}\ and\ \citenamefont
  {Winitzki}(2007)}]{Mukhanov:2007zz}%
  \BibitemOpen
  \bibfield  {author} {\bibinfo {author} {\bibfnamefont {V.}~\bibnamefont
  {Mukhanov}}\ and\ \bibinfo {author} {\bibfnamefont {S.}~\bibnamefont
  {Winitzki}},\ }\href@noop {} {\emph {\bibinfo {title} {{Introduction to
  quantum effects in gravity}}}}\ (\bibinfo  {publisher} {Cambridge University
  Press},\ \bibinfo {year} {2007})\BibitemShut {NoStop}%
\bibitem [{\citenamefont {Landau}\ and\ \citenamefont {Lifshitz}(1976)}]{LL}%
  \BibitemOpen
  \bibfield  {author} {\bibinfo {author} {\bibfnamefont {L.}~\bibnamefont
  {Landau}}\ and\ \bibinfo {author} {\bibfnamefont {S.}~\bibnamefont
  {Lifshitz}, \bibfnamefont {E.}},\ }\href@noop {} {\emph {\bibinfo {title}
  {{Mechanics, Vol.1, pp. 80-84}}}}\ (\bibinfo  {publisher}
  {Butterworth-Heinemann},\ \bibinfo {year} {1976})\BibitemShut {NoStop}%
\bibitem [{\citenamefont {Marinov}\ and\ \citenamefont
  {Popov}(1977)}]{Marinov:1977gq}%
  \BibitemOpen
  \bibfield  {author} {\bibinfo {author} {\bibfnamefont {M.~S.}\ \bibnamefont
  {Marinov}}\ and\ \bibinfo {author} {\bibfnamefont {V.~S.}\ \bibnamefont
  {Popov}},\ }\bibfield  {title} {\bibinfo {title} {{Electron-Positron Pair
  Creation from Vacuum Induced by Variable Electric Field}},\ }\href
  {https://doi.org/10.1002/prop.19770250111} {\bibfield  {journal} {\bibinfo
  {journal} {Fortsch. Phys.}\ }\textbf {\bibinfo {volume} {25}},\ \bibinfo
  {pages} {373} (\bibinfo {year} {1977})}\BibitemShut {NoStop}%
\bibitem [{\citenamefont {Brezin}\ and\ \citenamefont
  {Itzykson}(1970)}]{Brezin:1970xf}%
  \BibitemOpen
  \bibfield  {author} {\bibinfo {author} {\bibfnamefont {E.}~\bibnamefont
  {Brezin}}\ and\ \bibinfo {author} {\bibfnamefont {C.}~\bibnamefont
  {Itzykson}},\ }\bibfield  {title} {\bibinfo {title} {{Pair production in
  vacuum by an alternating field}},\ }\href
  {https://doi.org/10.1103/PhysRevD.2.1191} {\bibfield  {journal} {\bibinfo
  {journal} {Phys. Rev. D}\ }\textbf {\bibinfo {volume} {2}},\ \bibinfo {pages}
  {1191} (\bibinfo {year} {1970})}\BibitemShut {NoStop}%
\bibitem [{\citenamefont {Ringwald}(2001)}]{Ringwald:2001ib}%
  \BibitemOpen
  \bibfield  {author} {\bibinfo {author} {\bibfnamefont {A.}~\bibnamefont
  {Ringwald}},\ }\bibfield  {title} {\bibinfo {title} {{Pair production from
  vacuum at the focus of an X-ray free electron laser}},\ }\href
  {https://doi.org/10.1016/S0370-2693(01)00496-8} {\bibfield  {journal}
  {\bibinfo  {journal} {Phys. Lett. B}\ }\textbf {\bibinfo {volume} {510}},\
  \bibinfo {pages} {107} (\bibinfo {year} {2001})},\ \Eprint
  {https://arxiv.org/abs/hep-ph/0103185} {arXiv:hep-ph/0103185} \BibitemShut
  {NoStop}%
\bibitem [{\citenamefont {Schwinger}(1951)}]{Schwinger:1951nm}%
  \BibitemOpen
  \bibfield  {author} {\bibinfo {author} {\bibfnamefont {J.~S.}\ \bibnamefont
  {Schwinger}},\ }\bibfield  {title} {\bibinfo {title} {{On gauge invariance
  and vacuum polarization}},\ }\href {https://doi.org/10.1103/PhysRev.82.664}
  {\bibfield  {journal} {\bibinfo  {journal} {Phys. Rev.}\ }\textbf {\bibinfo
  {volume} {82}},\ \bibinfo {pages} {664} (\bibinfo {year} {1951})}\BibitemShut
  {NoStop}%
\bibitem [{\citenamefont {Coleman}\ and\ \citenamefont
  {Weinberg}(1973)}]{Coleman:1973jx}%
  \BibitemOpen
  \bibfield  {author} {\bibinfo {author} {\bibfnamefont {S.~R.}\ \bibnamefont
  {Coleman}}\ and\ \bibinfo {author} {\bibfnamefont {E.~J.}\ \bibnamefont
  {Weinberg}},\ }\bibfield  {title} {\bibinfo {title} {{Radiative Corrections
  as the Origin of Spontaneous Symmetry Breaking}},\ }\href
  {https://doi.org/10.1103/PhysRevD.7.1888} {\bibfield  {journal} {\bibinfo
  {journal} {Phys. Rev. D}\ }\textbf {\bibinfo {volume} {7}},\ \bibinfo {pages}
  {1888} (\bibinfo {year} {1973})}\BibitemShut {NoStop}%
\bibitem [{\citenamefont {Goldberg}(1990)}]{Goldberg:1990qk}%
  \BibitemOpen
  \bibfield  {author} {\bibinfo {author} {\bibfnamefont {H.}~\bibnamefont
  {Goldberg}},\ }\bibfield  {title} {\bibinfo {title} {{Breakdown of
  perturbation theory at tree level in theories with scalars}},\ }\href
  {https://doi.org/10.1016/0370-2693(90)90628-J} {\bibfield  {journal}
  {\bibinfo  {journal} {Phys. Lett. B}\ }\textbf {\bibinfo {volume} {246}},\
  \bibinfo {pages} {445} (\bibinfo {year} {1990})}\BibitemShut {NoStop}%
\bibitem [{\citenamefont {Cornwall}(1990)}]{Cornwall:1990hh}%
  \BibitemOpen
  \bibfield  {author} {\bibinfo {author} {\bibfnamefont {J.~M.}\ \bibnamefont
  {Cornwall}},\ }\bibfield  {title} {\bibinfo {title} {{On the High-energy
  Behavior of Weakly Coupled Gauge Theories}},\ }\href
  {https://doi.org/10.1016/0370-2693(90)90850-6} {\bibfield  {journal}
  {\bibinfo  {journal} {Phys. Lett. B}\ }\textbf {\bibinfo {volume} {243}},\
  \bibinfo {pages} {271} (\bibinfo {year} {1990})}\BibitemShut {NoStop}%
\bibitem [{\citenamefont {Flassig}\ \emph {et~al.}(2013)\citenamefont
  {Flassig}, \citenamefont {Pritzel},\ and\ \citenamefont
  {Wintergerst}}]{MacroQ1}%
  \BibitemOpen
  \bibfield  {author} {\bibinfo {author} {\bibfnamefont {D.}~\bibnamefont
  {Flassig}}, \bibinfo {author} {\bibfnamefont {A.}~\bibnamefont {Pritzel}},\
  and\ \bibinfo {author} {\bibfnamefont {N.}~\bibnamefont {Wintergerst}},\
  }\bibfield  {title} {\bibinfo {title} {{Black holes and quantumness on
  macroscopic scales}},\ }\href {https://doi.org/10.1103/PhysRevD.87.084007}
  {\bibfield  {journal} {\bibinfo  {journal} {Phys. Rev. D}\ }\textbf {\bibinfo
  {volume} {87}},\ \bibinfo {pages} {084007} (\bibinfo {year} {2013})},\
  \Eprint {https://arxiv.org/abs/1212.3344} {arXiv:1212.3344 [hep-th]}
  \BibitemShut {NoStop}%
\bibitem [{\citenamefont {Kovtun}\ and\ \citenamefont
  {Zantedeschi}(2020)}]{Kovtun:2020ndc}%
  \BibitemOpen
  \bibfield  {author} {\bibinfo {author} {\bibfnamefont {A.}~\bibnamefont
  {Kovtun}}\ and\ \bibinfo {author} {\bibfnamefont {M.}~\bibnamefont
  {Zantedeschi}},\ }\bibfield  {title} {\bibinfo {title} {{Breaking BEC}},\
  }\href {https://doi.org/10.1007/JHEP07(2020)212} {\bibfield  {journal}
  {\bibinfo  {journal} {JHEP}\ }\textbf {\bibinfo {volume} {07}},\ \bibinfo
  {pages} {212}},\ \Eprint {https://arxiv.org/abs/2003.10283} {arXiv:2003.10283
  [hep-th]} \BibitemShut {NoStop}%
\bibitem [{\citenamefont {'t~Hooft}(1987)}]{tHooft}%
  \BibitemOpen
  \bibfield  {author} {\bibinfo {author} {\bibfnamefont {G.}~\bibnamefont
  {'t~Hooft}},\ }\bibfield  {title} {\bibinfo {title} {{Graviton Dominance in
  Ultrahigh-Energy Scattering}},\ }\href
  {https://doi.org/10.1016/0370-2693(87)90159-6} {\bibfield  {journal}
  {\bibinfo  {journal} {Phys. Lett. B}\ }\textbf {\bibinfo {volume} {198}},\
  \bibinfo {pages} {61} (\bibinfo {year} {1987})}\BibitemShut {NoStop}%
\bibitem [{\citenamefont {Amati}\ \emph {et~al.}(1987)\citenamefont {Amati},
  \citenamefont {Ciafaloni},\ and\ \citenamefont {Veneziano}}]{Amati}%
  \BibitemOpen
  \bibfield  {author} {\bibinfo {author} {\bibfnamefont {D.}~\bibnamefont
  {Amati}}, \bibinfo {author} {\bibfnamefont {M.}~\bibnamefont {Ciafaloni}},\
  and\ \bibinfo {author} {\bibfnamefont {G.}~\bibnamefont {Veneziano}},\
  }\bibfield  {title} {\bibinfo {title} {{Superstring Collisions at Planckian
  Energies}},\ }\href {https://doi.org/10.1016/0370-2693(87)90346-7} {\bibfield
   {journal} {\bibinfo  {journal} {Phys. Lett. B}\ }\textbf {\bibinfo {volume}
  {197}},\ \bibinfo {pages} {81} (\bibinfo {year} {1987})}\BibitemShut
  {NoStop}%
\bibitem [{\citenamefont {Amati}\ \emph {et~al.}(1988)\citenamefont {Amati},
  \citenamefont {Ciafaloni},\ and\ \citenamefont {Veneziano}}]{AmatiStar}%
  \BibitemOpen
  \bibfield  {author} {\bibinfo {author} {\bibfnamefont {D.}~\bibnamefont
  {Amati}}, \bibinfo {author} {\bibfnamefont {M.}~\bibnamefont {Ciafaloni}},\
  and\ \bibinfo {author} {\bibfnamefont {G.}~\bibnamefont {Veneziano}},\
  }\bibfield  {title} {\bibinfo {title} {{Classical and Quantum Gravity Effects
  from Planckian Energy Superstring Collisions}},\ }\href
  {https://doi.org/10.1142/S0217751X88000710} {\bibfield  {journal} {\bibinfo
  {journal} {Int. J. Mod. Phys. A}\ }\textbf {\bibinfo {volume} {3}},\ \bibinfo
  {pages} {1615} (\bibinfo {year} {1988})}\BibitemShut {NoStop}%
\bibitem [{\citenamefont {Gross}\ and\ \citenamefont {Mende}(1988)}]{Gross}%
  \BibitemOpen
  \bibfield  {author} {\bibinfo {author} {\bibfnamefont {D.~J.}\ \bibnamefont
  {Gross}}\ and\ \bibinfo {author} {\bibfnamefont {P.~F.}\ \bibnamefont
  {Mende}},\ }\bibfield  {title} {\bibinfo {title} {{String Theory Beyond the
  Planck Scale}},\ }\href {https://doi.org/10.1016/0550-3213(88)90390-2}
  {\bibfield  {journal} {\bibinfo  {journal} {Nucl. Phys. B}\ }\textbf
  {\bibinfo {volume} {303}},\ \bibinfo {pages} {407} (\bibinfo {year}
  {1988})}\BibitemShut {NoStop}%
\bibitem [{\citenamefont {Gross}\ and\ \citenamefont
  {Mende}(1987)}]{GrossStar}%
  \BibitemOpen
  \bibfield  {author} {\bibinfo {author} {\bibfnamefont {D.~J.}\ \bibnamefont
  {Gross}}\ and\ \bibinfo {author} {\bibfnamefont {P.~F.}\ \bibnamefont
  {Mende}},\ }\bibfield  {title} {\bibinfo {title} {{The High-Energy Behavior
  of String Scattering Amplitudes}},\ }\href
  {https://doi.org/10.1016/0370-2693(87)90355-8} {\bibfield  {journal}
  {\bibinfo  {journal} {Phys. Lett. B}\ }\textbf {\bibinfo {volume} {197}},\
  \bibinfo {pages} {129} (\bibinfo {year} {1987})}\BibitemShut {NoStop}%
\bibitem [{\citenamefont {Dvali}\ and\ \citenamefont {Gomez}(2010)}]{selfC}%
  \BibitemOpen
  \bibfield  {author} {\bibinfo {author} {\bibfnamefont {G.}~\bibnamefont
  {Dvali}}\ and\ \bibinfo {author} {\bibfnamefont {C.}~\bibnamefont {Gomez}},\
  }\bibfield  {title} {\bibinfo {title} {{Self-Completeness of Einstein
  Gravity}},\ }\href@noop {} {\  (\bibinfo {year} {2010})},\ \Eprint
  {https://arxiv.org/abs/1005.3497} {arXiv:1005.3497 [hep-th]} \BibitemShut
  {NoStop}%
\bibitem [{\citenamefont {Dvali}\ and\ \citenamefont
  {Venugopalan}(2022)}]{Dvali:2021ooc}%
  \BibitemOpen
  \bibfield  {author} {\bibinfo {author} {\bibfnamefont {G.}~\bibnamefont
  {Dvali}}\ and\ \bibinfo {author} {\bibfnamefont {R.}~\bibnamefont
  {Venugopalan}},\ }\bibfield  {title} {\bibinfo {title} {{Classicalization and
  unitarization of wee partons in QCD and gravity: The CGC-black hole
  correspondence}},\ }\href {https://doi.org/10.1103/PhysRevD.105.056026}
  {\bibfield  {journal} {\bibinfo  {journal} {Phys. Rev. D}\ }\textbf {\bibinfo
  {volume} {105}},\ \bibinfo {pages} {056026} (\bibinfo {year} {2022})},\
  \Eprint {https://arxiv.org/abs/2106.11989} {arXiv:2106.11989 [hep-th]}
  \BibitemShut {NoStop}%
\bibitem [{\citenamefont {Gelis}\ \emph {et~al.}(2010)\citenamefont {Gelis},
  \citenamefont {Iancu}, \citenamefont {Jalilian-Marian},\ and\ \citenamefont
  {Venugopalan}}]{CGC}%
  \BibitemOpen
  \bibfield  {author} {\bibinfo {author} {\bibfnamefont {F.}~\bibnamefont
  {Gelis}}, \bibinfo {author} {\bibfnamefont {E.}~\bibnamefont {Iancu}},
  \bibinfo {author} {\bibfnamefont {J.}~\bibnamefont {Jalilian-Marian}},\ and\
  \bibinfo {author} {\bibfnamefont {R.}~\bibnamefont {Venugopalan}},\
  }\bibfield  {title} {\bibinfo {title} {{The Color Glass Condensate}},\ }\href
  {https://doi.org/10.1146/annurev.nucl.010909.083629} {\bibfield  {journal}
  {\bibinfo  {journal} {Ann. Rev. Nucl. Part. Sci.}\ }\textbf {\bibinfo
  {volume} {60}},\ \bibinfo {pages} {463} (\bibinfo {year} {2010})},\ \Eprint
  {https://arxiv.org/abs/1002.0333} {arXiv:1002.0333 [hep-ph]} \BibitemShut
  {NoStop}%
\bibitem [{\citenamefont {Gross}\ and\ \citenamefont
  {Neveu}(1974)}]{GrossNeveu}%
  \BibitemOpen
  \bibfield  {author} {\bibinfo {author} {\bibfnamefont {D.~J.}\ \bibnamefont
  {Gross}}\ and\ \bibinfo {author} {\bibfnamefont {A.}~\bibnamefont {Neveu}},\
  }\bibfield  {title} {\bibinfo {title} {{Dynamical Symmetry Breaking in
  Asymptotically Free Field Theories}},\ }\href
  {https://doi.org/10.1103/PhysRevD.10.3235} {\bibfield  {journal} {\bibinfo
  {journal} {Phys. Rev. D}\ }\textbf {\bibinfo {volume} {10}},\ \bibinfo
  {pages} {3235} (\bibinfo {year} {1974})}\BibitemShut {NoStop}%
\bibitem [{\citenamefont {Dvali}\ and\ \citenamefont
  {Sakhelashvili}(2022)}]{GNsaturons}%
  \BibitemOpen
  \bibfield  {author} {\bibinfo {author} {\bibfnamefont {G.}~\bibnamefont
  {Dvali}}\ and\ \bibinfo {author} {\bibfnamefont {O.}~\bibnamefont
  {Sakhelashvili}},\ }\bibfield  {title} {\bibinfo {title} {{Black-hole-like
  saturons in Gross-Neveu}},\ }\href
  {https://doi.org/10.1103/PhysRevD.105.065014} {\bibfield  {journal} {\bibinfo
   {journal} {Phys. Rev. D}\ }\textbf {\bibinfo {volume} {105}},\ \bibinfo
  {pages} {065014} (\bibinfo {year} {2022})},\ \Eprint
  {https://arxiv.org/abs/2111.03620} {arXiv:2111.03620 [hep-th]} \BibitemShut
  {NoStop}%
\bibitem [{\citenamefont {Dvali}(2018{\natexlab{b}})}]{Dvali:2018xpy}%
  \BibitemOpen
  \bibfield  {author} {\bibinfo {author} {\bibfnamefont {G.}~\bibnamefont
  {Dvali}},\ }\bibfield  {title} {\bibinfo {title} {{A Microscopic Model of
  Holography: Survival by the Burden of Memory}},\ }\href@noop {} {\  (\bibinfo
  {year} {2018}{\natexlab{b}})},\ \Eprint {https://arxiv.org/abs/1810.02336}
  {arXiv:1810.02336 [hep-th]} \BibitemShut {NoStop}%
%%CITATION = ARXIV:1810.02336;%%
\bibitem [{\citenamefont {Ellis}(2017)}]{Ellis:2016jkw}%
  \BibitemOpen
  \bibfield  {author} {\bibinfo {author} {\bibfnamefont {J.}~\bibnamefont
  {Ellis}},\ }\bibfield  {title} {\bibinfo {title} {{TikZ-Feynman: Feynman
  diagrams with TikZ}},\ }\href {https://doi.org/10.1016/j.cpc.2016.08.019}
  {\bibfield  {journal} {\bibinfo  {journal} {Comput. Phys. Commun.}\ }\textbf
  {\bibinfo {volume} {210}},\ \bibinfo {pages} {103} (\bibinfo {year}
  {2017})},\ \Eprint {https://arxiv.org/abs/1601.05437} {arXiv:1601.05437
  [hep-ph]} \BibitemShut {NoStop}%
\end{thebibliography}

\end{document}